\documentclass[aps,pre, superscriptaddress]{revtex4-1}
\usepackage[dvips]{graphicx}
\usepackage{amsmath}
\usepackage{amssymb}
\usepackage{natbib}
\usepackage{color}
\usepackage{epsfig}
\usepackage{subfigure}
\usepackage{setspace}
\begin{document}
\title{Mean-field density of states of a small-world model and a jammed soft spheres model}

 \author{Mario Pernici}
\email{mario.pernici@mi.infn.it}
\affiliation{ Istituto Nazionale di Fisica Nucleare, Sezione di Milano,\\ 16 Via Celoria, 20133 Milano, Italy}
\begin{abstract}
We consider a class of random block matrix models in $d$ dimensions,
$d \ge 1$, motivated by the study of the vibrational density of states (DOS)
of  soft spheres near the isostatic point.
The contact networks of average degree $Z = z_0 + \zeta$ are represented by 
random $z_0$-regular graphs (only the circle graph in $d=1$ with $z_0=2$)
to which Erd\"os-Renyi graphs having a small
average degree $\zeta$ are superimposed.

In the case $d=1$, for $\zeta$ small the shifted Kesten-McKay 
DOS with parameter $Z$ is a mean-field solution for the DOS.
Numerical simulations in the $z_0=2$ model, which is
the $k=1$ Newman-Watts small-world model, and in the $z_0=3$ model
lead us to conjecture that for $\zeta \to 0$ the cumulative function of the DOS 
converges uniformly
to that of the shifted Kesten-McKay DOS, in an interval $[0, \omega_0]$, with $\omega_0 < \sqrt{z_0-1} + 1$.

For $2 \le d \le 4$, we introduce a cutoff parameter $K_d \le 0.5$ 
modeling sphere repulsion.
The case $K_d=0$ is the random
elastic network case, with the DOS close to the Marchenko-Pastur DOS
with parameter $t=\frac{Z}{d}$.
For $K_d$ large the DOS is close for small $\omega$
to the shifted Kesten-McKay DOS with parameter $t=\frac{Z}{d}$; in the isostatic case the DOS has around 
$\omega=0$ the expected plateau.
The boson peak frequency in $d=3$ with $K_3$ large is close to
the one found in molecular dynamics simulations for $Z=7$ and $8$.

 \end{abstract}
\maketitle

\section{Introduction}
The spectral distribution of the vibrational modes of disordered solids
is not yet completely understood, in particular the origin of the sharp
increase of the vibrational density of states (DOS) $D(\omega)$ at small
frequency, around the frequency identified as the
    "boson peak" in $\frac{D(\omega)}{\omega^{d-1}}$, which gives a
comparison between $D(\omega)$ and the Debye law $D(\omega) \sim \omega^{d-1}$,
where $d$ is the dimension.

Modeling a disordered solid with frictionless soft spheres with repulsive
finite range potentials, in \cite{ohern} it has been found in computer
simulations that at the jamming threshold
$D(\omega)$ has a plateau around
zero frequency, and that the network of contacts has average 
coordination degree $Z$ equal to $Z_c=2d$.
This excess of low energy modes is related to the condition of isostaticity
    introduced by Maxwell \cite{maxwell}, balancing the number of degrees of freedom of 
the particles with the constraints between them.
In the hyperstatic region, in which $Z > 2d$ and 
$\delta Z \equiv Z - 2d$ is small, there is a boson peak.
In \cite{wnw} it is argued, based on the Maxwell condition, that the sharp increase
    in $D(\omega)$ is a common feature of weakly-connected amorphous solids.

In crystals a sharp increase in the DOS occurs at a van Hove singularity,
which occurs at high frequencies, related to the short-range properties of
the crystal. The boson peak seems to have some characteristics typical
    of long-range properties (occurring at small frequency) and 
short-range properties.
    It has been suggested that the boson peak
    is a shifted van Hove singularity \cite{tar}.

In \cite{MZ} a local inversion-symmetry breaking parameter has been found
to be relevant for the boson peak; it is sensitive to the angular correlation
between bonds connecting spheres.

In \cite{ML} it has been observed that random matrix models with translational
invariance give a possible simple model of the boson peak.

Random block matrix models on random regular graphs have been
studied in \cite{pari1}, to give a simple mean-field approximation
of the random network of contacts, in which the versors pointing
from a node to another are independent and randomly uniformly distributed
in $d$ dimensions.
In that paper it has been shown that, for $d\to\infty$,
the spectral distribution of the Hessian matrix for an elastic network on random regular graphs
    is the Marchenko-Pastur (MP) distribution.
    This model has been studied in
\cite{pari1, benetti} in $d=3$ using the cavity method and exact
diagonalization, together with a class of models ${\cal G}_{(z_0,\zeta)}$
in which the random graphs with average degree $Z = z_0 + \zeta$
are formed superimposing Erd\"os-Renyi random graphs of degree $\zeta$
to random regular graphs of degree $z_0$.
In \cite{benetti} it has been remarked that, due to the tree-like nature of
the random regular graphs, Goldstone modes are absent, so there are
no phonons, but only strongly-scattered modes.
They find that in the isostatic case the DOS has a peak
for zero frequency, possibly a logarithmic divergence, instead of having a plateau.
In the hyperstatic case ${\cal G}_{(7,0)}$ in $d=3$ their numerical
simulations show that there is a quasi-gap with $D(\omega) \sim \omega^4$
for $\omega\to 0$, predicted in \cite{gurev} and found in molecular
dynamics simulations in \cite{ler, miz, kap}.
 
The method of the spectral distribution moments for random block matrix 
models  on random Erd\"os-Renyi graphs has been studied in \cite{CKMZ}, 
where it was conjectured that the spectral distribution
in the limit $d\to \infty$, with $\frac{Z}{d}$ fixed, is the
MP distribution.
This conjecture has been proved in \cite{CP}.

Another class of models characterized by the interplay between short-range
properties and long-range properties are the small-world models \cite{ws, nw}.
In particular  a mean-field analytic form for the distribution
of path lengths has been found \cite{nmw} in the $k=1$ Newman-Watts 
small-world model \cite{nw}, a ring with a small number of random shortcuts;
this mean-field solution tends to be exact in the limit of a large number
of nodes and few shortcuts.
Various properties of the vibrational spectrum of this model have been
studied:
the effective medium approximation (EMA) spectrum density for this model 
for small $\omega$ is given in \cite{monasson};
the lowest non-zero eigenvalue of the laplacian matrix, known as the
algebraic connectivity \cite{fiedler} has been studied in \cite{olf,gu};
a lower bound on it is given in \cite{gu}.
The laplacian spectral distribution for this small-world model is discussed in 
\cite{olf}.

The spectral distribution of the adjacency matrix of an uniformly distributed random $t$-regular graph has been found in \cite{kesten,MK}; shifting it one obtains the spectral distribution of the corresponding laplacian matrix,
and hence the DOS of a random regular elastic network in $d=1$.
We will indicate in the following with $D_K(\omega; t)$ 
the shifted Kesten-McKay DOS (SKM DOS) with parameter $t$.

In this paper we study the DOS of the laplacian random block matrix for
${\cal G}_{(z_0,\zeta)}$ models in various dimensions; for $d=1$ it is
the laplacian matrix of a random graph, for $d > 1$ the blocks are, as in
\cite{benetti,CKMZ}, $d$-dimensional projectors defined by the random $d$-dimensional
contact versor between two sites.

In section II we review the random elastic network model. 
Using \cite{CP} we give
a new proof that in the case of a random regular graph network of degree
$Z$ in $d$ dimensions, in the limit $d\to \infty$ and $t=\frac{Z}{d}$
fixed one obtains the MP distribution with parameter $t$.

In section III we argue that in the ${\cal G}_{(z_0,\zeta)}$ model in $d=1$,
with average degree $t=z_0+\zeta$, for $\zeta$ small
the density of state is well approximated by the SKM DOS with parameter $t$, 
by giving a mean-field argument, based on the corresponding 
derivation for the random regular graphs \cite{wanless}.
Numerical simulations in the cases $z_0=2$ and $3$
lead us to conjecture that for $\zeta \to 0$ the cumulative function
of the DOS tends uniformly to the one of the SKM DOS,
in an interval $[0, \omega_0]$, with $\omega_0 < \sqrt{z_0-1} + 1$.
The ${\cal G}_{(2,\zeta)}$ model is the $k=1$ Newman-Watts small-world model.

In section IV we introduce a model for jammed soft spheres; it differs
from the random network model by the introduction of
a cutoff parameter $K_d$ representing sphere repulsion in an equilibrium
configuration: 
in its absence ($K_d=0$) one has the random network
model studied in \cite{pari1, benetti}, in which the contact versors
are independent random variables, uniformly distributed in $d$ dimensions;
for $0 < K_d \le 0.5$ 
two random versors $v,w$ representing the directions of the contacts of
a sphere with two other spheres satisfy the bound
$v\cdot w \le 1-K_d$.

Numerical simulations in $d=2,3,4$ indicate that, in the isostatic case
${\cal G}_{(2d,0)}$
the peak of $D(\omega)$ in $\omega=0$, present for $K_d = 0$,
becomes with the increase of $K_d$ the expected plateau.

We find that in the ${\cal G}_{(z_0,\zeta)}$
models in  $1 \le d \le 4$, with large cutoff $K_d$ for $2 \le d \le 4$,
$D(\omega)$ is well approximated for $\omega < 1.5$ by the
SKM DOS $D_K(\omega; \frac{Z}{d})$, apart from
a long tail around $\omega=0$ in $d=1$.
This is to be contrasted with the $K_d = 0$ case, with $d=2,3,4$,
in which the DOS is much closer to the MP DOS than to the SKM DOS.
Therefore the parameter $K_d$ allows an interpolation between 
a case close the $d=1$ solution (SKM)
and a case close to the $d=\infty$ solution (MP). 
In the $d=3$, $Z=7$ and $8$ hyperstatic cases, the boson peak frequencies
corresponding to the SKM DOS are close to the values found
in molecular dynamics simulation in \cite{MZ}.

For $K_d$ large, the quantity $\frac{D(\omega)}{\omega^2}$, whose peak is the boson
peak in the $d=3$ model,
is well approximated by $\frac{D_K(\omega; \frac{Z}{d})}{\omega^2}$.

For $2 \le d \le 4$ we find that there is a gap in $D(\omega)$
close to the one in $D_K(\omega; \frac{Z}{d})$; the quasi-gap
found in \cite{benetti} in $d=3$ with $D(\omega) \approx \omega^4$
for $\omega \to 0$ in the hyperstatic case $Z=7$ 
becomes a gap in presence of a large cutoff $K_3$.

In section II we review the network model for soft-spheres near
jamming.
In section III we study the $d=1$ models, including the $k=1$ Newman-Watts small-world model.
In section IV we study the soft-sphere model with angular cutoff.

\section{Random block matrix model for a random elastic network}
Consider a system of $N$ soft spheres in $d$ dimensions, with centers
in $r_i$, $i=1,\cdots,N$. We assume that the spheres interact with
a short-range repulsive central potential. 
Following \cite{pari1,benetti},
to an equilibrium configuration of spheres a network of contacts is associated,
where an edge $(i,j)$ is present if the sphere in $r_i$ is in contact
with the sphere in $r_j$. Let $v_{i,j} = r_j - r_i$;
$\hat v_{i,j}$ is the corresponding versor.
In a random elastic network the contact versors are random independent
uniformly distributed $d$-dimensional Gaussian versors.

Define $X_{i,j} = |\hat v_{i,j}><\hat v_{i,j}|$.
In the elastic approximation, in which displacements $r_i \to r_i + \delta_i$ 
are kept only to quadratic order, the energy of the system is, neglecting the
"initial-stress" contribution,
\begin{equation}
E = \sum_{i,j} \delta_i\cdot M_{i,j}\cdot \delta_j
\end{equation}
where $M_{i,j}$ is the Hessian,
\begin{eqnarray}
M_{i,j} &=& -\alpha_{i,j} X_{i,j}, \qquad i \neq j = 1,\cdots,N \nonumber \\
M_{i,i} &=& \sum_{j \neq i} \alpha_{i,j} X_{i,j}
\label{eqM}
\end{eqnarray}
with $\alpha_{i,j}$ element of the adjacency matrix of the network graph.
$M$ has eigenvalues $\lambda = \omega^2$, where $\omega$ is the frequency.

The density of states is
\begin{equation}
D(\omega) = 2\omega \rho(\omega^2)
\label{Drho}
\end{equation}
where $\rho(\lambda)$ is the spectral density.

The Maxwell isostatic condition is $Z_c = 2d$;
$Z$ is the average coordination degree. 

In \cite{ohern} it has been observed that
jamming occurs at the isostatic point; 
the density of states $D(\omega)$ has a plateau around $\omega=0$.
Close to the jamming point they found with numerical simulations that
$ Z - Z_c \propto (\phi -\phi_c)^\frac{1}{2}$, 
$\phi$ being the packing fraction and $\phi_c$ the jamming
packing fraction. Therefore $Z$ plays a similar role to the
packing fraction.

In a physical network of contacts
the coordination numbers should be close to the average degree $Z$.
In \cite{pari1} the network of contacts has been chosen to be random
regular graphs.

The exact DOS for these models for arbitrary $d$ is not known; the $d=1$
 random regular graph case is obtained from the Kesten-McKay distribution;
 it is reviewed in section III.

The $d \to \infty$ random regular graph model, with $t=\frac{Z}{d}$ fixed,
gives the MP distribution \cite{pari1}.
Let us give a new proof of this, using \cite{CP}.

The contributions to the spectral moments are given, in the limit
$N\to \infty$, by tree walks. 
In the case of random regular graphs of degree $Z$, the walks are
on the Bethe lattice of degree $Z$.

Consider the unoriented graph on which the walk is embedded.
It has the root of degree $\delta_0$ and the other nodes $i$ of degree
$\delta_i$.

Starting from the root, the first edge can be chosen in $Z$ ways;
when the path returns to the root, an edge different from the one 
previously taken can be chosen in $Z-1$ ways, and so on, so one gets
a factor $Z^{\underline {\delta_0}}$,
where $n^{\underline i}$ is the falling factorial.
When the path arrives at the node $i$ different from the root, there
is at that node at least an edge already visited, so that
one gets a factor $(Z-1)^{\underline{\delta_i-1}}$.

Therefore there is a combinatorial factor associated to the walk
on the Bethe lattice
\begin{equation}
Z^{\underline {\delta_0}} \prod_i (Z-1)^{\underline{\delta_i-1}}
\label{zfact}
\end{equation}

In the limit $d\to \infty$ with $t=\frac{Z}{d}$ fixed,
the $Z-i$ factors can be approximated with $Z$ at leading order,
so one gets a factor $Z^E$, where $E$ is the number of edges
in the tree associated to the walk.
Then one gets the same power of $Z$ that one gets from the same
walk in the Erd\"os-Renyi model, in the limit $N\to \infty$.
In that case the $Z$ factors come from the probability distribution
of the adjacency matrix weights.

The average on the blocks $X_{i,j}$ associated to the steps of the
walk is the same in the Erd\"os-Renyi and in the random regular graph cases;
in the limit $d\to \infty$ it gives to leading order a factor $d^{-E}$;
putting the two factors together one gets a factor $t^E$.

Therefore in the limit $N\to \infty$, $d\to \infty$ with $t=\frac{Z}{d}$ fixed
one gets the same spectral moments in the two models.
Since in \cite{CP} it has been proved that in the case of Erd\"os-Renyi
graphs the spectral distribution is the MP distribution, the same holds
for the case of random regular graphs.

To the MP distribution corresponds the density of states

\begin{eqnarray}
D_M(\omega; t) &=& \frac{1}{2\pi\omega}\sqrt{(b^2-\omega^2)(\omega^2-a^2)}\nonumber \\
a &=& \sqrt{t} - \sqrt{2}\,; \qquad b = \sqrt{t} + \sqrt{2}
\label{densmp}
\end{eqnarray}
for $a \le \omega \le b$ and zero otherwise.

In \cite{CKMZ} it has been observed that the random block matrix model
with Erd\"os-Renyi contact network
interpolates between the $d=1$ model with coordination degree $t$ and the 
$d\to \infty$ model with coordination degree $t=\frac{Z}{d}$.
In \cite{CP} it has been proved that the non-crossing contributions
to the spectral moments depend only on $t$, so they are common to the
models with different $d$, but same $t=\frac{Z}{d}$. 
The crossing contributions depend on $t$ and $d$; they vanish
in the $d\to \infty$ limit. 
In \cite{CKMZ} simulations show that, for $d > 1$, the spectral distribution
is fairly well approximated by the one for $d\to \infty$, the MP distribution 
with parameter $t$, so the crossing contributions seem to have a small effect.

The $d=1$ spectral distribution differs markedly from the MP DOS, 
due to a series of spikes.  Its analytical form is unknown.

The Kesten-McKay distribution has been derived in \cite{kesten,MK}
for $t$-regular random graphs, hence with $t$ integer.
In the next section we will discuss $d=1$ models, which have as mean-field
DOS the SKM DOS with $t$ real.

The uniformity of the DOS of the Erd\"os-Renyi models with different
$d$ and $t=\frac{Z}{d}$ fixed, leads us to examine with simulations
whether the same happens in the case of random regular graphs,
and to see whether in the latter case the DOS is closer to the $d=1$ SKM DOS 
or the $d\to \infty$ Marchenko-Pastur DOS, both with parameter $t$.
In section IV we will see that in the elastic network model the DOS
is close to the MP DOS with parameter $t$, while in a soft sphere model with large
angular cutoff, modeling the repulsion between spheres in an equilibrium
configuration, the DOS is close to the SKM DOS with parameter $t$.

\section{$d=1$ models}

\subsection{Kesten-McKay distribution and ${\cal G}_{(z_0,\zeta)}$ models}
The Kesten-McKay distribution \cite{kesten, MK} is the spectral distribution 
of the Adjacency matrix $A$ of random uniformly distributed regular graphs of 
degree $t$, in the limit of number of nodes going to infinity.

Let us review here a derivation of this distribution \cite{wanless}.

In the limit of number of nodes $N$ going to infinity, the moment
of the spectral distribution $\mu_n = \frac{1}{N} \langle tr A^n \rangle$ 
is the number
of walks of length $n$ on a Bethe lattice of degree $t$.

Consider an infinite rooted tree with root of degree $s$ and all the
other nodes of degree $t$.
Let $T_s^{(t)}(x)$ be the generating function of the 
number of tree walks of length $n$ starting at the root; 
$T_s^{(t)}(0) = 1$.
A non-trivial walk starting at the root can go from the root to a
neighbor node in $s$ ways; then it can take a walk not returning
to the root; then it returns to the root. From there it can take another
tree walk.
One has therefore
\begin{equation}
T_s^{(t)} - 1 = x^2 s T_{t-1}^{(t)} T_s^{(t)}
\label{Ts}
\end{equation}
Taking the case $s=t-1$ in this equation, one obtains a quadratic equation,
whose root is determined by the condition $T_{t-1}^{(t)}(0) = 1$.
Substituting this in Eq. (\ref{Ts}), in the case $s=t$ one gets
\begin{equation}
T_t^{(t)}(x) = \frac{t-2-t\sqrt{1-4x^2(t-1)}}{2(x^2t^2-1)}
\end{equation}
The resolvent of the adjacency matrix is 
\begin{equation}
r_A(z) = \sum_{n\ge 0} \mu_n z^{-n-1} = \frac{1}{z}T_t^{(t)}(\frac{1}{z})
\end{equation}
From
\begin{equation}
\rho_A(x) = -\frac{1}{\pi} \lim_{\epsilon \to 0^+} \texttt{Im} \, r_A(x+i\epsilon)
\end{equation} 
one gets the spectral density of the adjacency matrix
\begin{equation}
\rho_A(z) = \frac{t}{2\pi}\frac{\sqrt{4(t-1)-z^2}}{t^2-z^2}
\label{rhokk}
\end{equation}
for $|z| \le 2\sqrt{t-1}$ and $0$ otherwise.

The Laplacian matrix differs from $-A$ only by $t$ times the unit matrix,
so that its spectral distribution is obtained shifting by $t$ the one
in Eq.(\ref{rhokk}).
The density of states is related to the Laplacian spectral density by
$D(\omega) = 2\omega \rho(\omega^2)$, $z=\omega^2$
so that the corresponding SKM density of states is
\begin{equation}
D_K(\omega; t) =
\frac{t\omega}{\pi} \frac{\sqrt{4(t-1)-(\omega^2-t)^2}}{t^2-(\omega^2-t)^2}
\label{bpk}
\end{equation}
for $\sqrt{t-1} - 1 < \omega < \sqrt{t-1} + 1$, zero otherwise.

For $t \ge 3$ this is the density of states for random $t$-regular graphs.
For $t=2$ it is the density of states for the ring
\begin{equation}
D_K(\omega; 2) = \frac{2}{\pi \sqrt{4-\omega^2}}
\label{bpkt2}
\end{equation}

While the SKM distribution holds exactly for random regular
graphs, we argue that for a class of tree-like random graphs with average degree
$t$ real it gives a good approximation to the density of states.
We consider a class of models ${\cal G}_{(z_0,\zeta)}$, with average degree 
$t = z_0 + \zeta$, in which the random graphs are obtained from a regular
graph (a ring for $z_0=2$, a randomly distributed $z_0$-regular graph for $z_0 \ge 3$),
to which a Erd\"os-Renyi graph of small average degree $\zeta$ is superimposed.
The degree of each vertex $i$ of the graph is $z_i = z_0 + \zeta_i$,
where $\zeta_i$ is a Poisson random variable having mean $\zeta$.
The degree distribution of each vertex is 
$p_k = \zeta^{k-z_0}\frac{e^{-\zeta}}{(k-z_0)!}$ for $k \ge z_0$, zero otherwise \cite{benetti}.

Let us consider, similarly to the case of random regular graphs,
the generating function $T_s^{(t)}$ of the average number of tree walks of
length $n$,
starting from the root with average degree $s$, the other nodes with
average degree $t$; thus $s$ and $t$ are now real quantities.

A non-trivial walk starting at the root can go from the root to a
neighbor node in $s$ ways; then it can take a walk not returning
to the root; then it returns to the root. From there it can take another
tree walk.

In a mean-field approximation one obtains again Eq.(\ref{Ts}),
where now $s$ and $t$ are average degrees,
and from it the spectral density for the Adjacency matrix Eq.(\ref{rhokk}).
For $\zeta$ small, the degrees are close to $t$, so approximatively the
Laplacian matrix is $t I - A$, so that one obtains approximatively the
density of states  $D_K(\omega; t)$ in Eq.(\ref{bpk}),
where $t$ is the average degree, with $t = z_0 + \zeta$.
As $\zeta$ approaches zero, this approximation can be expected
to become more accurate.

Due to the fluctuations in numerical simulations for $D(\omega)$, 
it is convenient to use the cumulative function
\begin{equation}
\Phi(\omega) = \int_0^\omega d \omega D(\omega)
\end{equation}
and similarly $\Phi_K(\omega)$ and $\Phi_M(\omega)$ respectively
for the SKM and the MP DOS.

\subsection{Density of states for the $k=1$ Newman-Watts small-world model}
In the case $z_0=2$, $t = 2 + \zeta$, the random graphs are defined as a circle graph 
with $N$ nodes, to which a Erd\"os-Renyi graph of average degree $\zeta$ 
is superimposed.

This model belongs to the class of Newman-Watts models, in which the random graph
is formed by a ring on $N$ nodes, each node being connected to $2k$ nearest
neighbors; an Erd\"os-Renyi graph is superimposed to it \cite{nw}.
In the $k=1$ Newman-Watts model, in \cite{nmw} they have
computed an analytic expression for the distribution of the distance
between nodes, for $\zeta$ small. They obtain it considering
a mean-field approximation, in which the average of the distributions
on the random graphs is considered. For $\zeta \to 0$
this approximation becomes exact.

We find a similar behavior for the DOS,
where $D_K(\omega; 2+\zeta)$ is the mean-field approximation.

A discussion of the spectrum of the Laplacian for the $k$-small-world models
is done in \cite{monasson}, with emphasis on the simulations in the $k=3$
small-world model. The $k=1$ small-world model is briefly mentioned there,
giving the EMA approximation
\begin{equation}
D_E(\omega; \zeta) = \frac{\sqrt{8\omega^2 - \zeta^2}}{2\pi \omega}
\label{ema}
\end{equation}
to which corresponds the cumulative function $\Phi_E(\omega)$.

We make numerical simulations to see how the DOS tends to the SKM DOS
for $\zeta \to 0$.

In the simulations in this section we used $200$
random graphs with $3000$ nodes, unless specified.
The eigenvalues of the laplacian matrix are computed using in Sagemath
\cite{sage} the function  Numpy.linalg.eigvalsh.

\begin{figure*}[h]
\begin{center}
\epsfig{file=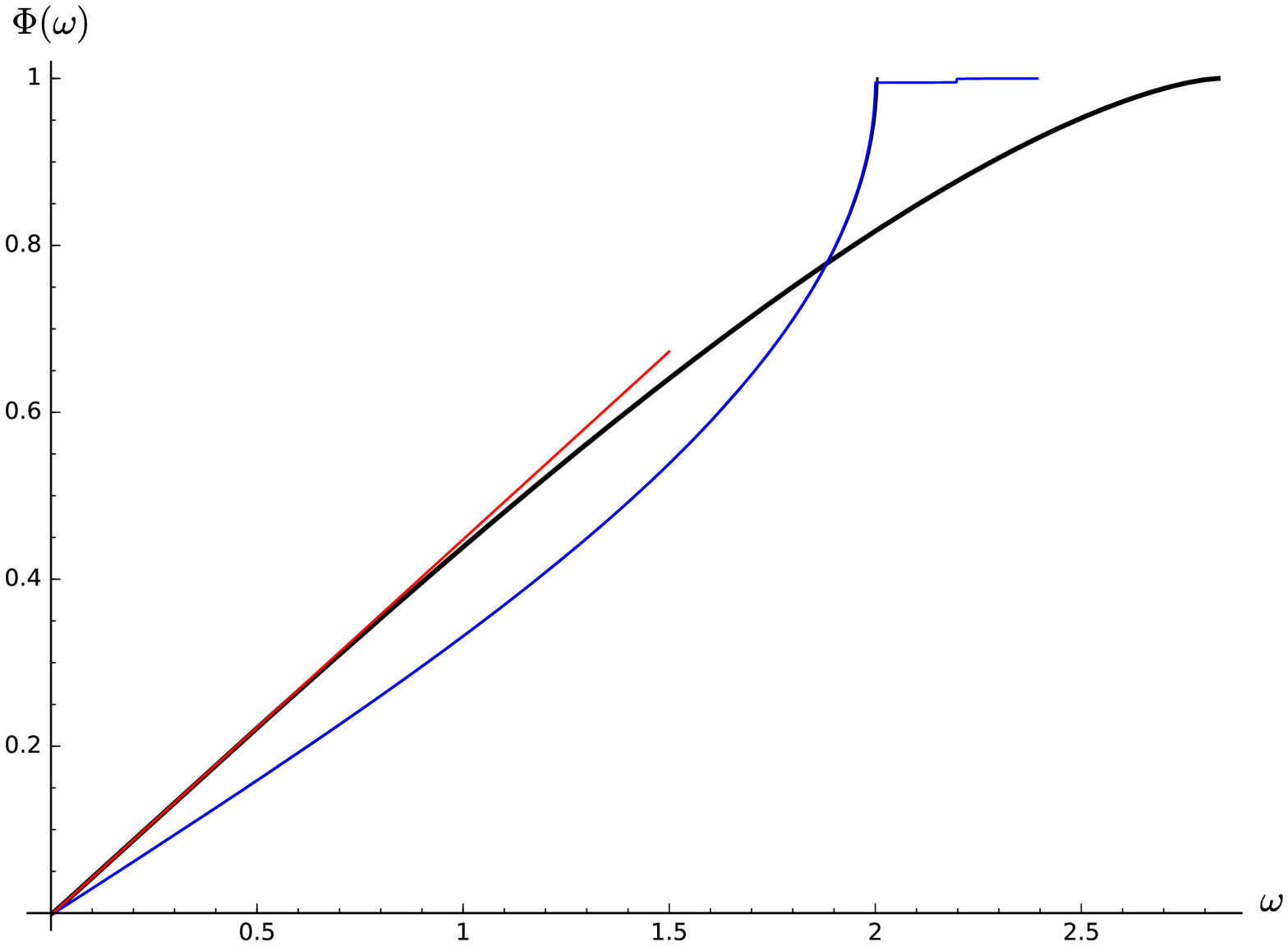, width=7.00cm}\quad
\epsfig{file=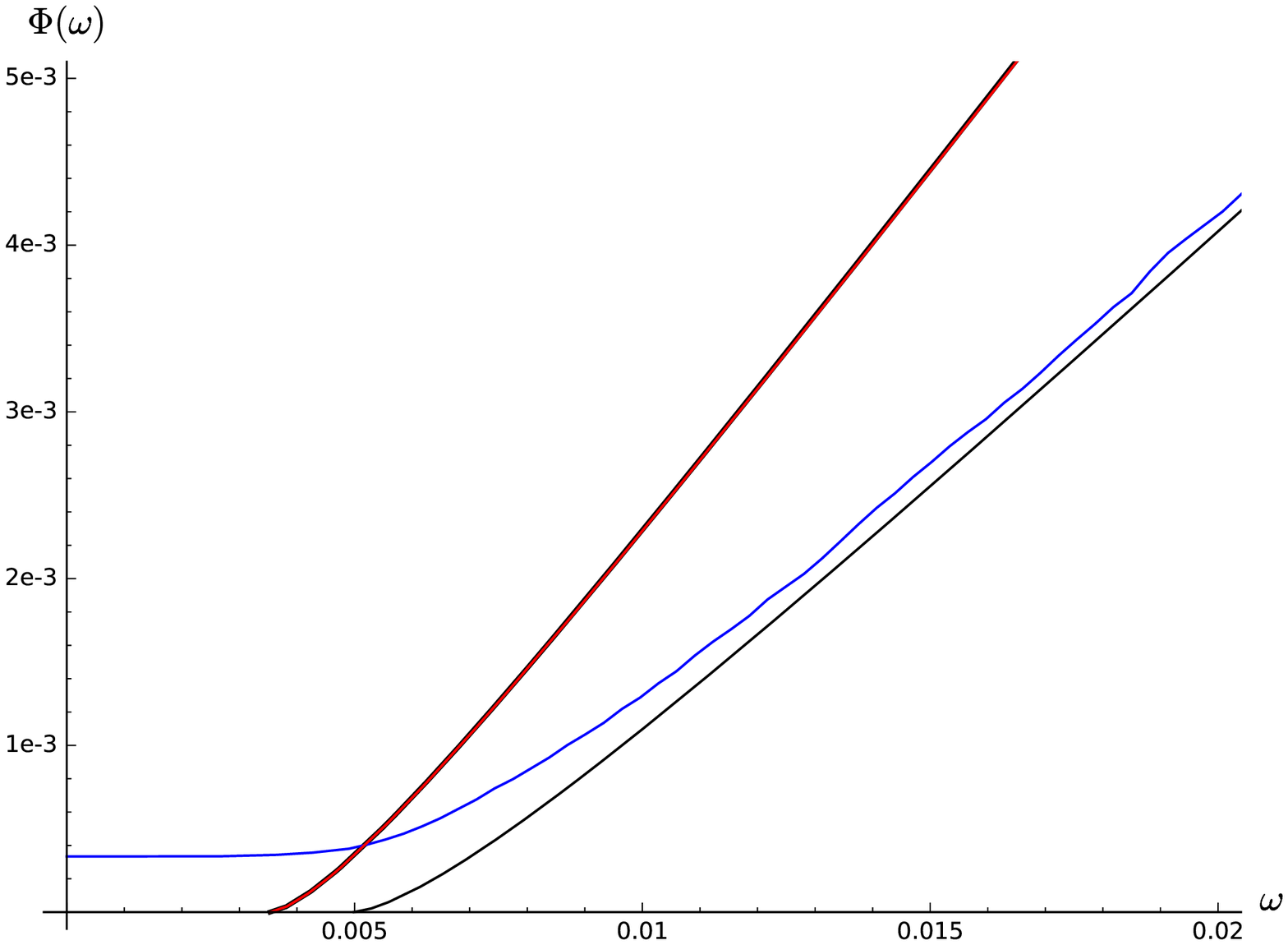, width=7.00cm}\quad
\caption{In the left hand figure $\Phi(\omega)$ for $t=2.01$ is plotted
in blue, the SKM in black (thin line) and the MP DOS as a thick black line.
The EMA DOS is plotted for $\omega < 1.5$, in red.
In the right hand figure, they are plotted for small $\omega$; the EMA DOS
is very close to the MP DOS.
}
\label{Figphi201}
\end{center}
\end{figure*}

In Fig. \ref{Figphi201} the cumulative function is given for the case
${\cal G}_{(2,0.01)}$; in the left hand figure are plotted $\Phi$,
$\Phi_K$, $\Phi_M$ and $\Phi_E$.
$\Phi$ has a high frequency (HF) tail, with $\omega > \sqrt{t-1} + 1$, beyond
the support of the SKM DOS.
In $\omega=2.1976(3)$ the DOS has a spike (see Fig. \ref{Fig201} and its note),
giving a dent in $\Phi$, visible in Fig. \ref{Figphi201}.
In the right hand figure the cumulative functions for $\zeta=0.01$ are given for
small $\omega$; $\Phi$ starts with the zero-mode contribution.

\begin{figure*}[h]
\begin{center}
\epsfig{file=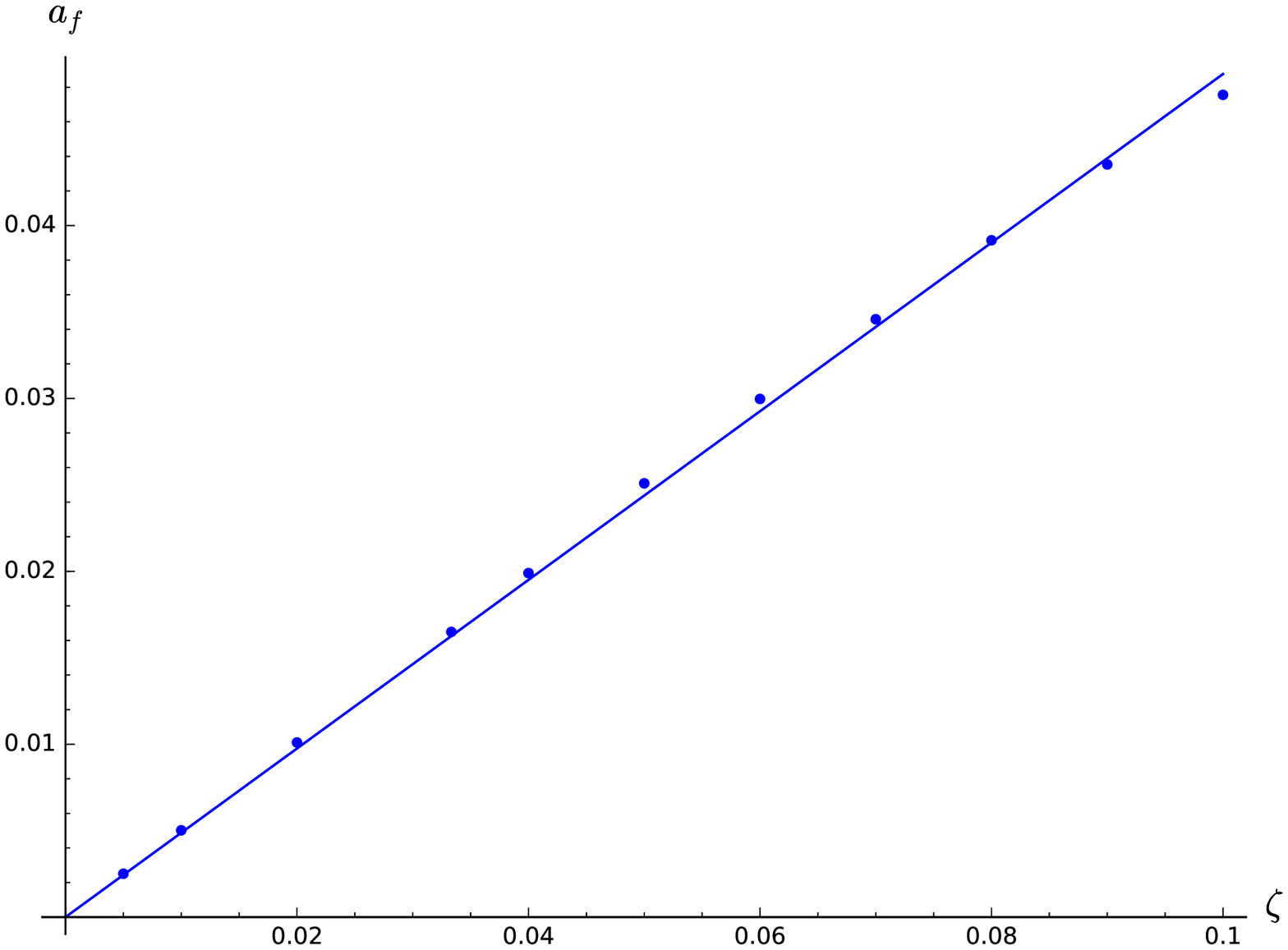, width=5.00cm} \quad
\epsfig{file=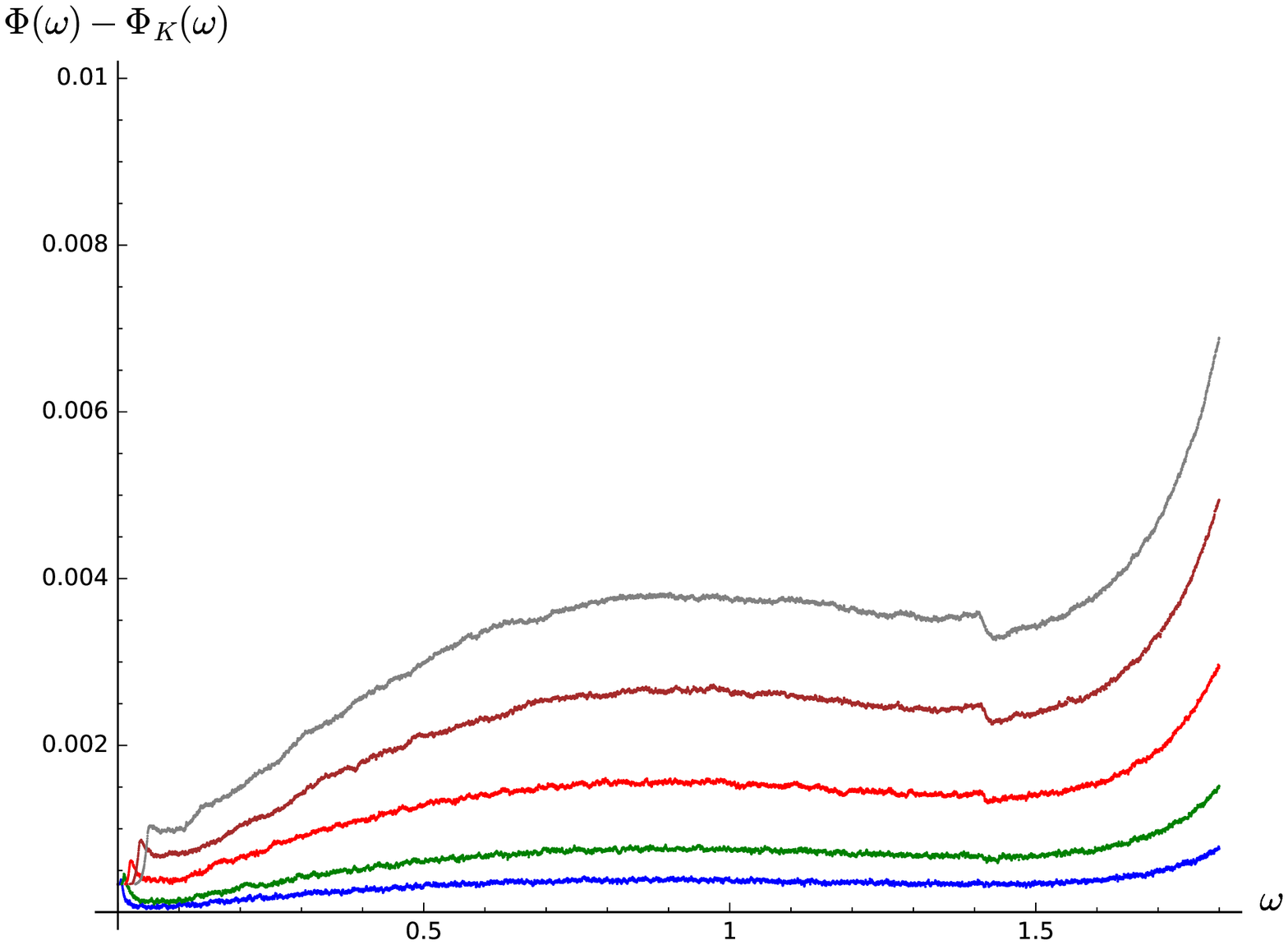, width=5.00cm}\quad
\epsfig{file=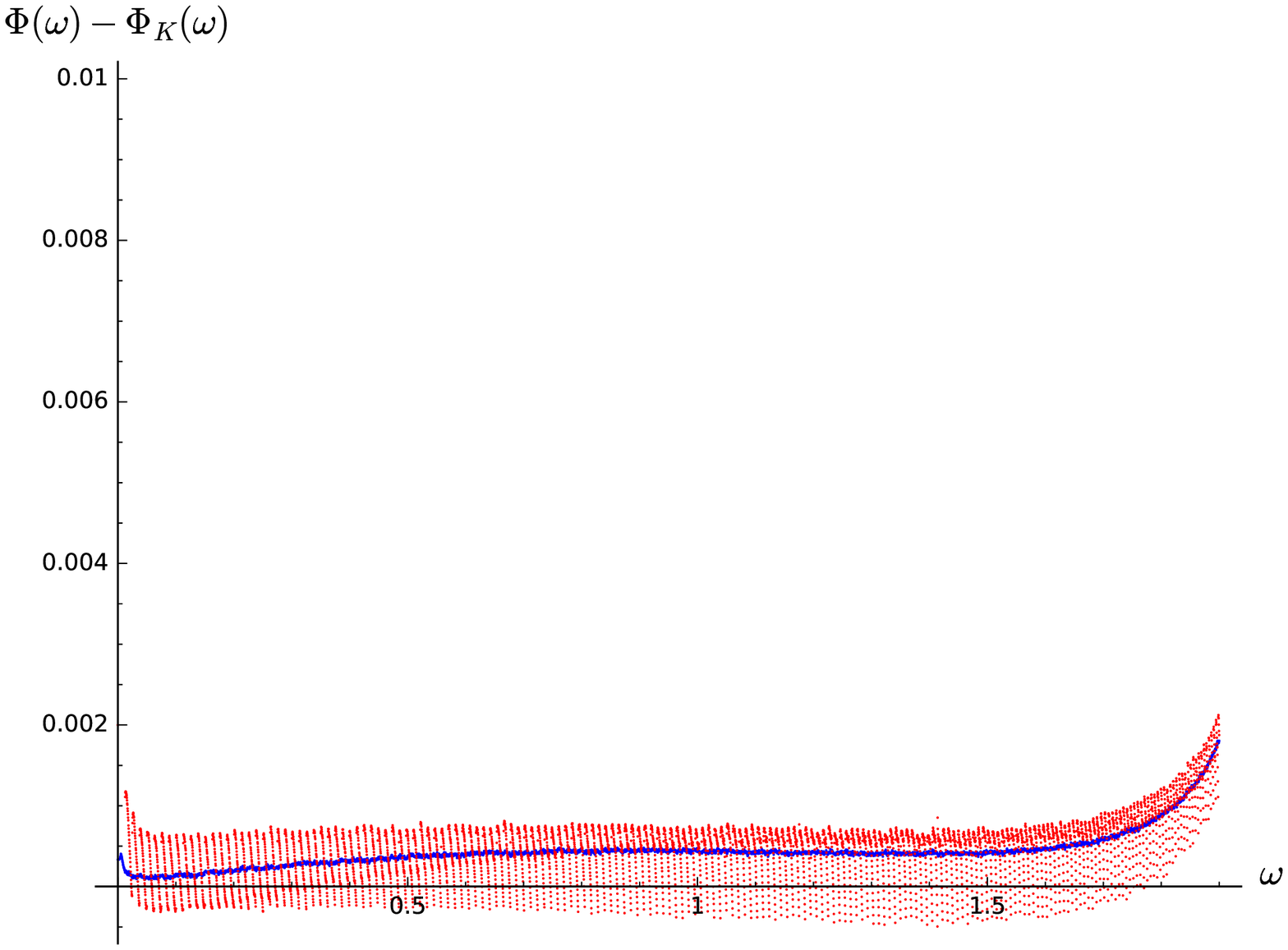, width=5.00cm}
\caption{In the case ${\cal G}_{(2,\zeta)}$,
the left hand figure shows the area  $a_f$ under $D(\omega)$ in the HF tail
for various values of $\zeta$ between $0.005$ and $0.1$, for $N=3000$.
The center figure shows $\Phi(\omega) - \Phi_K(\omega)$ for $N=3000$,
    $\zeta=0.01$(blue), $0.02$(green), $0.04$(red), $0.07$(brown) and $0.1$(gray) for $\omega < 1.8$.
The right hand figure shows $\Phi(\omega) - \Phi_K(\omega)$ for
    $\zeta=0.01$, $N=500$(red) and $N=3000$ (blue).
}
\label{Figpvlia}
\end{center}
\end{figure*}

In the left hand figure in Fig. \ref{Figpvlia} are given the areas under
$D(\omega)$
in the HF tails, for a few values of $\zeta$ between $0.005$
and $0.1$; they fit with $0.50 \zeta$, so they should go to zero
approximately at this rate.

The DOS has a van Hove peak tending to $2$ for $\zeta \to 0$: 
its frequency is $1.9987(3)$ for $\zeta=0.1$, $1.9998(3)$ for $\zeta=0.01$;
the uncertainty is due to the fact that we used bins of size
$\Delta \omega=10^{-3.5}$.

In the center figure in Fig. \ref{Figpvlia}, $\Phi(\omega) - \Phi_K(\omega)$
is plotted for a few values of $\zeta$ between $\zeta=0.01$ and $0.1$.
For $\omega > 1.8$ there is a sharp increase in this difference,
due to the fact that the DOS has the van Hove peak slightly 
less than $2$, while the SKM DOS has a van Hove peak at slightly less than
$\sqrt{t-1} + 1$. 

$\Phi$ starts at $\omega=0$ with the zero mode;
$\Phi(0)$ goes to zero for $N\to\infty$; with the increase of $N$, $\Phi$
has fewer fluctuations; in the right hand side figure in Fig. \ref{Figpvlia}
we give $\Phi(\omega) - \Phi_K(\omega)$
for $\zeta=0.01$, $N=500$ and $N=3000$.

The area of the high-frequency tail
does not change with $N$; e.g. for $\zeta=0.01$ it is equal to $0.005$ for
$N=500$, $1000$ and $3000$.

In the center figure in Fig. \ref{Fig201} for $t=2.01$, and Fig. \ref{Fig233} 
for $t=7/3$, one can see 
a difference also for $\omega$ small, where $D_K(\omega; t)$ is zero for
$\omega \le \sqrt{t-1} - 1$, while $D(\omega)$
has a tail going much closer to zero.
The smallest nonzero eigenvalue of the laplacian is the algebraic connectivity 
$\lambda_2=\omega_2^2$; it has been studied in \cite{olf,gu};
in \cite{gu} it has been shown to go slowly to zero for $N\to \infty$.

These simulations lead us to conjecture that the DOS tends to the SKM DOS 
for $\zeta \to 0$ in the following way:
for $N\to \infty$, $\Phi$ 
converges uniformly as $\zeta \to 0$ to $\Phi_K$ for $\omega \le \omega_0$, with $\omega_0 < 2$;
the area of the high-frequency tail, with $\omega > \sqrt{1+\zeta} + 1$,
goes to zero linearly for $\zeta \to 0$.
In the interval $[\omega_0, \sqrt{1+\zeta} + 1]$ there are the
van Hove peak of the SKM DOS and the one of the DOS, tending to each
other for $\zeta \to 0$.

\subsection{${\cal G}_{(3,\zeta)}$ model}
In the ${\cal G}_{(3,\zeta)}$ model the random graphs are formed
by random $3$-regular graphs, to which a Erd\"os-Renyi graph of average 
degree $\zeta$ is superimposed.

With respect to the $k=1$ Newman-Watts model, the algebraic connectivity $\lambda_2$
is larger;
in fact in a graph in the class ${\cal G}_{(z_0,\zeta)}$ one can expect that 
$\lambda_2 = \omega_2^2$ is larger than for a random $z_0$-regular graph,
for which the SKM DOS has lowest non-zero
frequency $\omega_2 = \sqrt{z_0-1} - 1$, hence $\sqrt{2} - 1$ for $z_0 = 3$.

\begin{figure*}[h]
\begin{center}
\epsfig{file=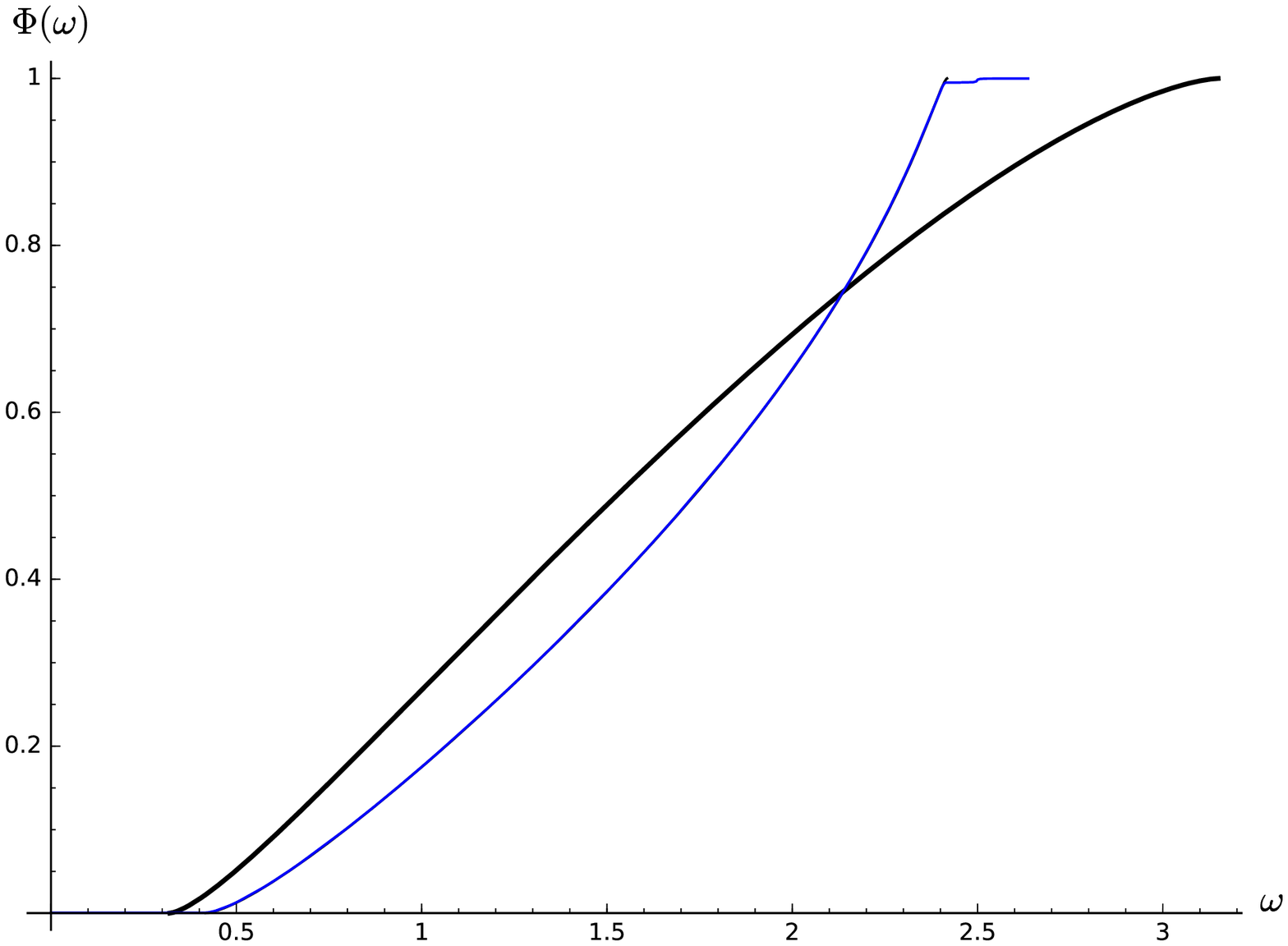, width=7.00cm}\quad
\epsfig{file=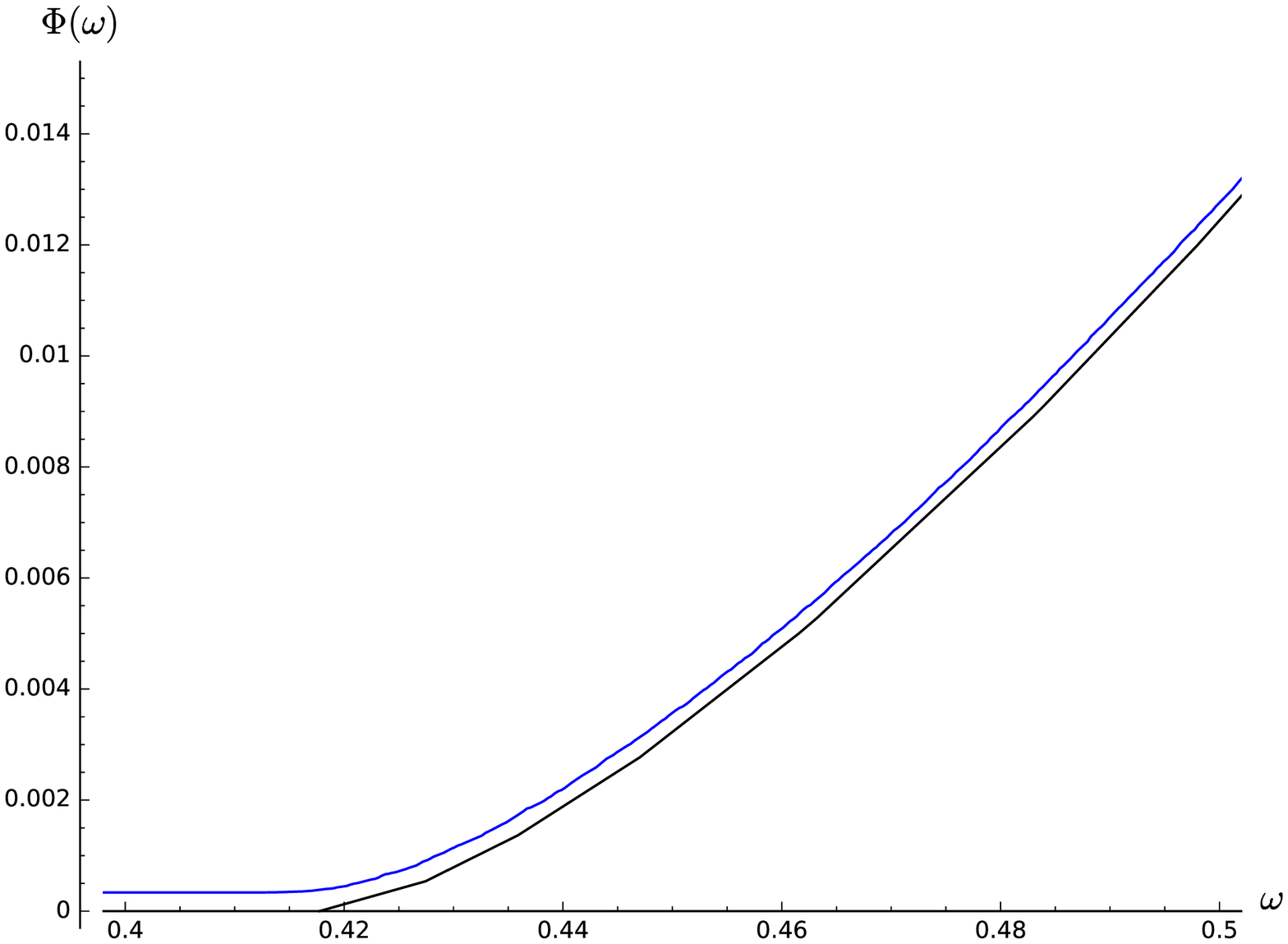, width=7.00cm}\quad
\caption{In the left hand figure are given the
cumulative functions for the case ${\cal G}_{(3,0.01)}$:
 $\Phi(\omega)$ (blue), 
$\Phi_K(\omega)$ (thin black),
$\Phi_M(\omega)$ (thick black).
In the right hand figure $\Phi(\omega)$ and 
$\Phi_K(\omega)$ are plotted between $\omega=0.4$ and $\omega=0.5$.
}
\label{Figphi301}
\end{center}
\end{figure*}

In the left hand figure in Fig. \ref{Figphi301} the cumulative function
is given for $\zeta=0.01$; in the right hand figure it is given near $\omega_2$.

\begin{figure*}[h]
\begin{center}
\epsfig{file=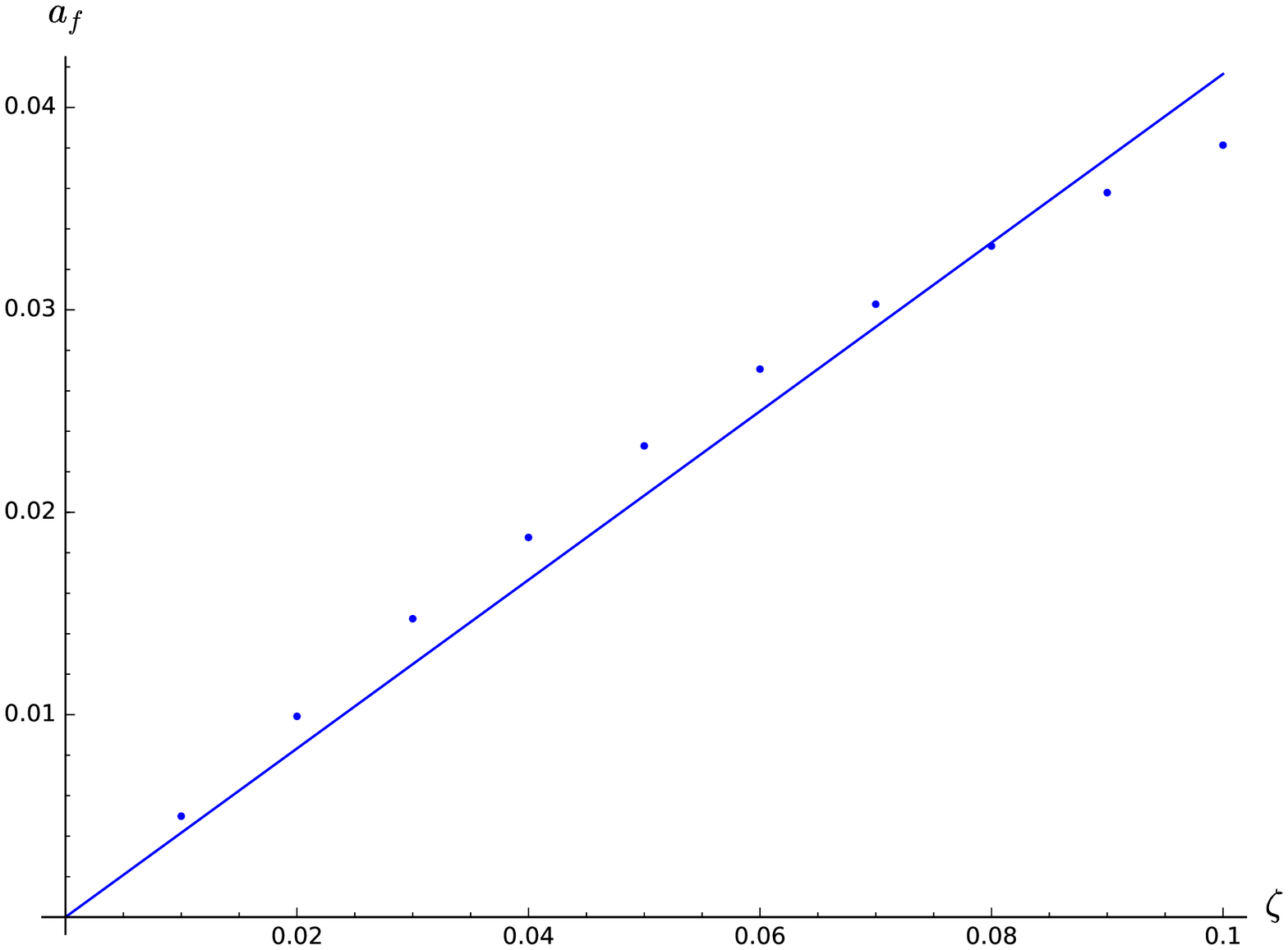, width=7.00cm}\quad
\epsfig{file=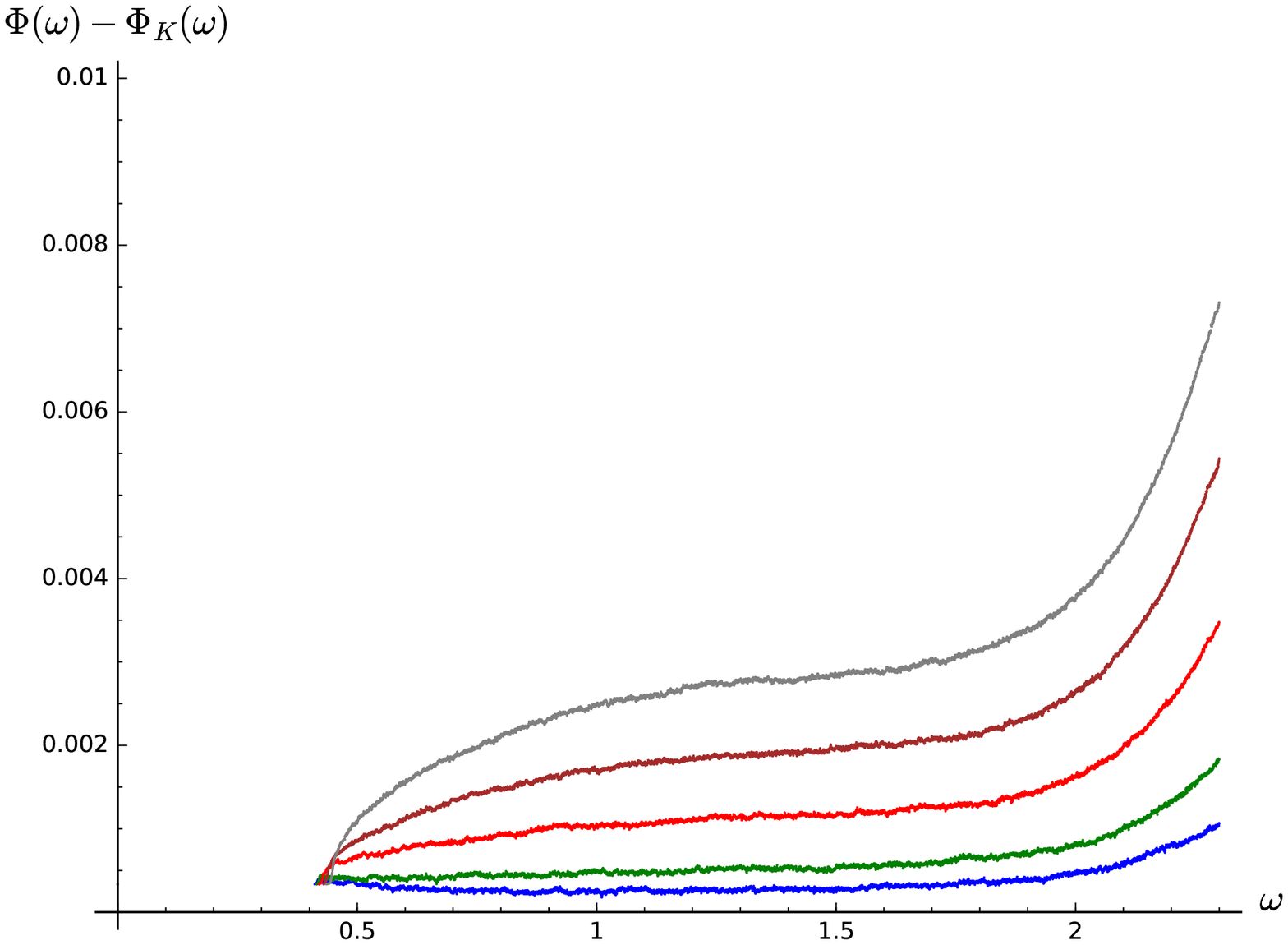, width=7.00cm}
\caption{In the case ${\cal G}_{(3,\zeta)}$,
the left hand figure shows the area  $a_f$ under of $D(\omega)$ in the HF tail
for various values of $\zeta$ between $0.01$ and $0.1$.
The right hand figure shows $\Phi(\omega) - \Phi_K(\omega)$ for 
$\omega < 2.3$, for
    $\zeta=0.01$(blue), $0.02$(green), $0.04$(red), $0.07$(brown) and $0.1$(gray).
}
\label{Figpvliaz3}
\end{center}
\end{figure*}

In the left hand figure in Fig. \ref{Figpvliaz3} are given the area under 
$D(\omega)$ in the HF region, i.e.  $1 - \Phi(\sqrt{t-1} + 1)$, for
various values of $\zeta$ between $0.01$ and $0.1$; a fit suggests
that this area goes to zero as $0.4 \zeta$ for $\zeta \to 0$.

In the right hand figure in Fig. \ref{Figpvliaz3},
$\Phi(\omega) - \Phi_K(\omega)$ is plotted for $\omega < 2.3$
for a few values of $\zeta$ between $0.01$ and $0.1$; this figure indicates
that for $\omega \le \omega_0$, where $\omega_0 < \sqrt{z_0-1} + 1$,
for $\zeta \to 0$ and $N\to \infty$ the difference
$\Phi(\omega) - \Phi_K(\omega)$ tends to zero.

Finally both the frequencies of van Hove peak of the SKM DOS and of
the one of the DOS tend to $\omega=1+\sqrt{2}$ for $\zeta \to 0$.

We conjecture that for $N\to\infty$, as $\zeta \to 0$ the cumulative function  
$\Phi$ converges uniformly to $\Phi_K$ in an interval
$[0, \omega_0]$, with $\omega_0 < \sqrt{z_0 - 1} + 1$; that the integral
of $D$ over the high frequency region $\omega > \sqrt{z_0 + \zeta - 1} + 1$
goes to $0$ in this limit. The frequency of the van Hove peak of the DOS tends to
$\sqrt{z_0 - 1} + 1$.

\clearpage

\section{A random block matrix model for jamming spheres with angular cutoff.}
The random elastic network model for jamming soft spheres introduced 
in \cite{pari1, benetti} and reviewed in section II does not take into 
account the repulsion between spheres in an equilibrium configuration.

\begin{figure*}[h]
\begin{center}
\epsfig{file=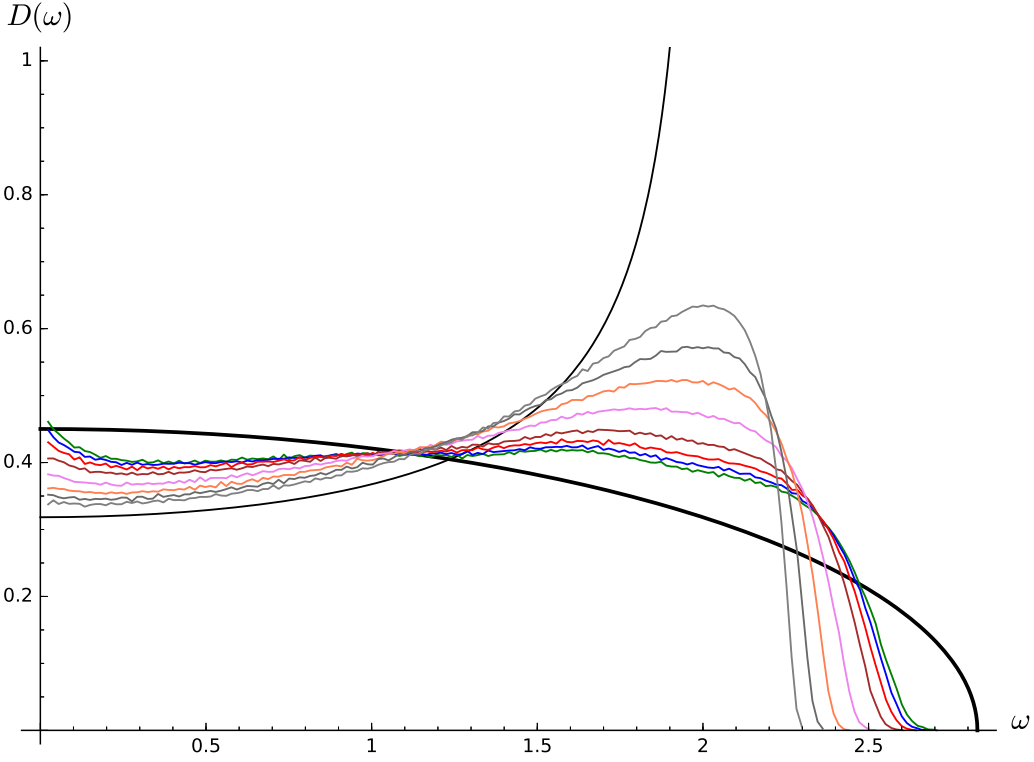, width=14.00cm} 
\caption{The figure shows $D(\omega)$  
in $d=3$ in the isostatic case $Z=6$.
Simulations are with $K_3=0$ (green), $K_3=0.02$(blue), $K_3=0.05$(red), 
$K_3=0.1$(brown), $K_3=0.2$(violet), $K_3=0.3$(coral),
$K_3=0.4$(dimgray), $K_3=0.5$(gray).
The smooth black curves are
the SKM DOS (thin curve) and the MP DOS (thick curve).
}
\label{Figiso}
\end{center}
\end{figure*}

In \cite{pari1, benetti} simulations were made for the random block matrix 
model for a random elastic network in $d=3$ using random
regular graphs. The resulting density of states was found to have, in the
isostatic case, a peak in $\omega=0$, instead of the expected plateau; 
in \cite{benetti} it is remarked that this peak might be
a logarithmic singularity. In the left hand figure in Fig. \ref{Figiso} this 
density of states is represented by a blue
line; it is fairly close to the MP DOS (the thick
black line) but, unlike it, it is not flat at $\omega=0$.

In an equilibrium configuration
spheres are kept apart by the repulsive force.
Let the sphere in $r_i$ be in contact with the spheres in $r_j$ and
$r_k$. Due to the repulsion potential spheres in contact are assumed
to be roughly at the same distance. To model this in the random block
matrix model, we add a parameter
$K_d$  to the random block matrix model, and the constraints
\begin{equation}
\hat v_{i,j}\cdot \hat v_{i,k} \le 1-K_d
\label{vect}
\end{equation}
for all $i,j,k$ such that $\alpha_{i,j}\alpha_{i,k} = 1$.
For $K_d=0$ one recovers the random network model considered previously.
For three hard spheres touching each other one has $K_d = \frac{1}{2}$,
so it is natural to assume $K_d \le \frac{1}{2}$.

For $K_d \neq 0$ the contact versors are not anymore independent
random versors uniformly distributed in $d$ dimensions.
We follow the following algorithm to choose the random versors.
In the simulations, to determine random contact versors satisfying
this cutoff constraint, we placed random vectors satisfying
Eq.(\ref{vect}) with a small cutoff, then we used a greedy algorithm
to get random vectors with the required cutoff. We accept the result
if the sum, over all sampled networks, of the violations to the required cutoff is less than $0.1$.

All the simulations in $d=3$ are made with 200 random graphs with $1000$
nodes.

In Fig. \ref{Figiso} there is the DOS in the isostatic $d=3$ case,
with a few values of $K_3$ from $0$ to $0.5$.
For $K_3=0.5$ there is a plateau around $\omega=0$; this DOS
is close to the SKM DOS with parameter $t=2$ (the ring DOS
Eq. (\ref{bpkt2})) for $\omega < 1.5$. 
The density of states with large $K_3$ has a  plateau for $\omega$ small;
with the decrease of $K_3$ a peak appears in $\omega=0$;
the plateau following the peak is also higher than in the
molecular dynamics simulations in \cite{ohern}.
The height of the plateau around $\omega=0$ depends on $K_3$; for 
$K_3 \simeq 0.3$ it fits with its height in molecular dynamics simulations.

Around $\omega=2$ there is a van Hove peak, which increases in height 
with $K_3$;
the SKM DOS has a much higher peak, which is not drawn in the figure.

\begin{figure*}[h]
\begin{center}
\epsfig{file=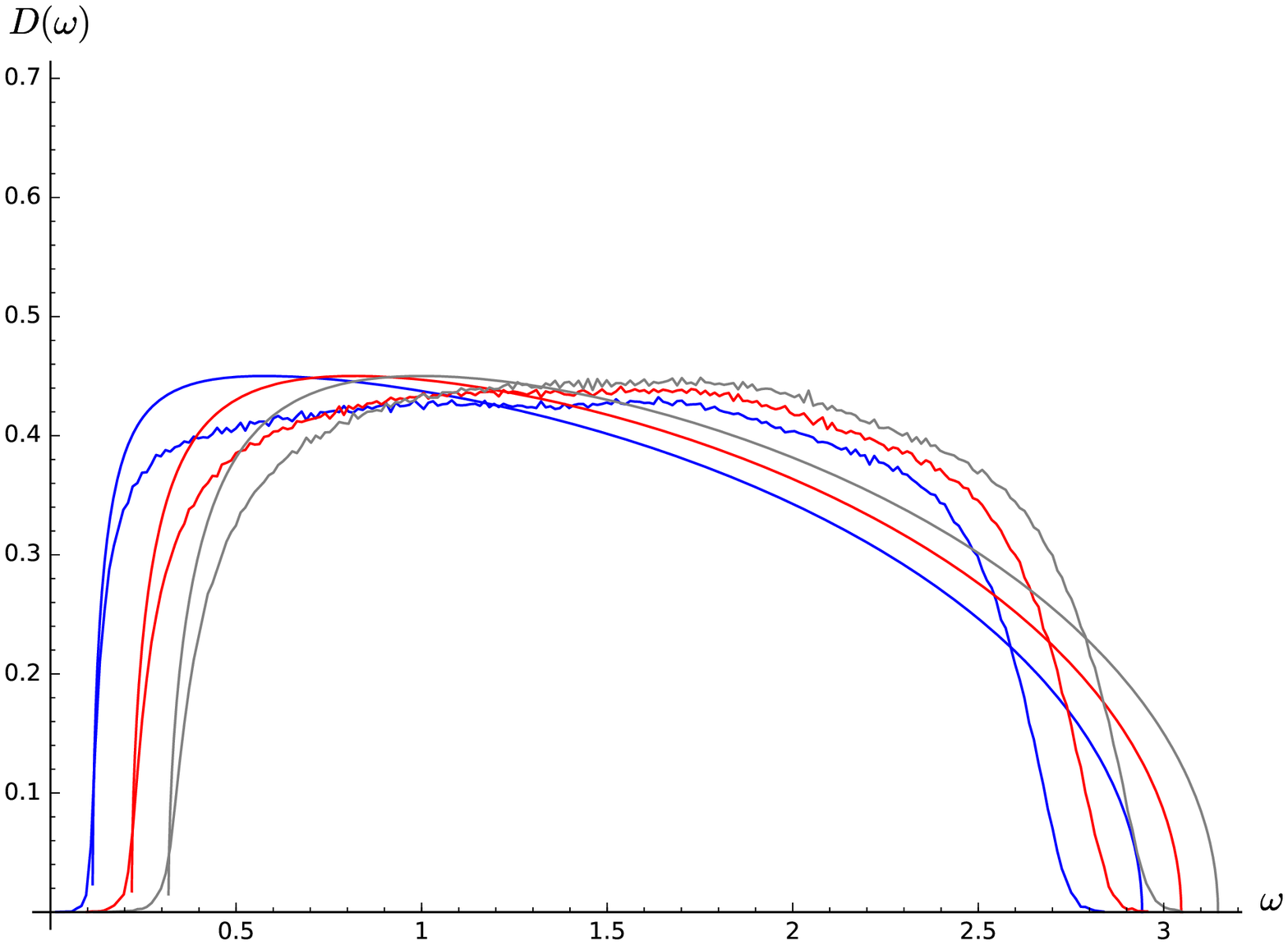, width=6.00cm} \qquad \epsfig{file=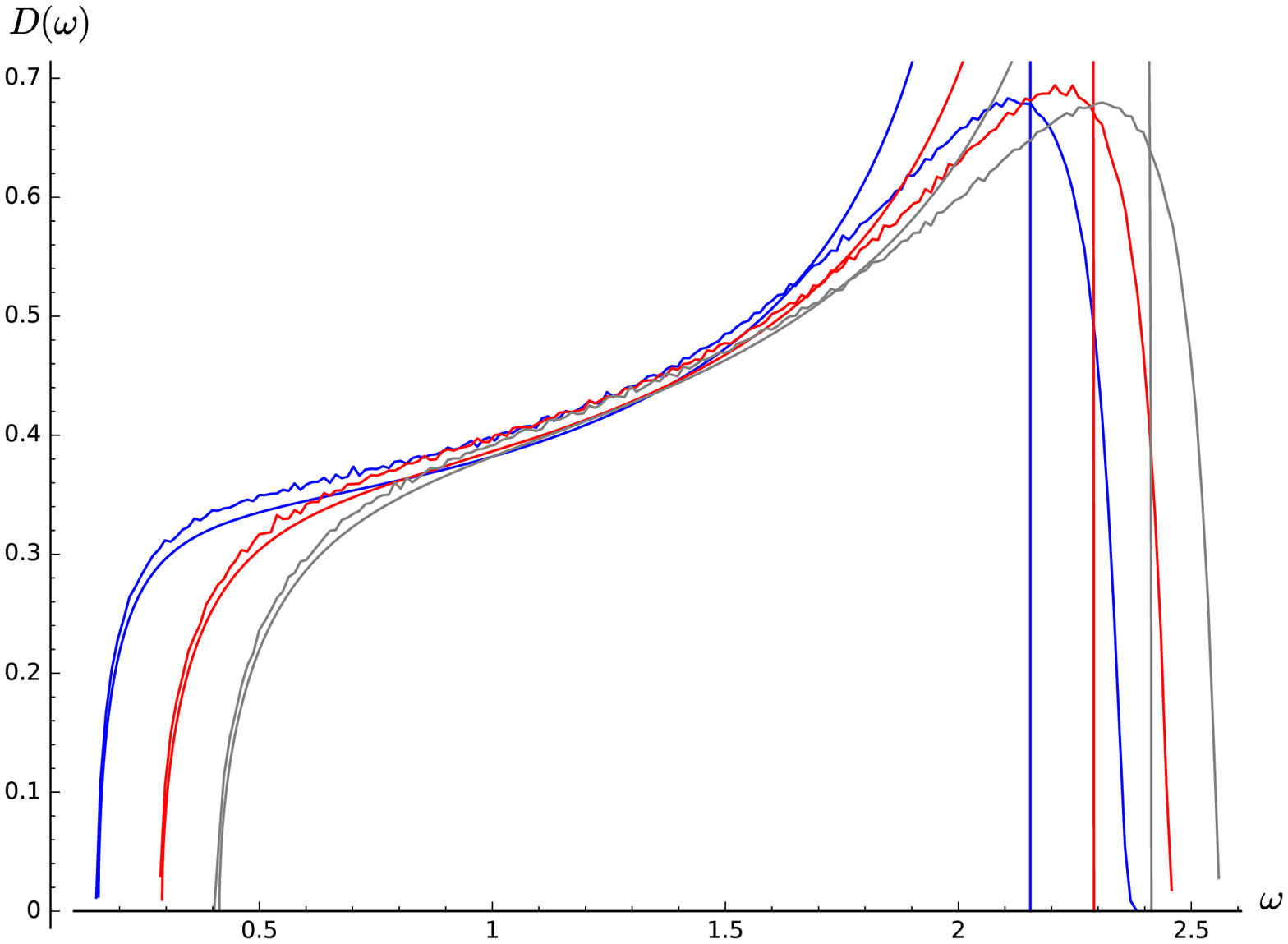, width=6.00cm} \qquad \epsfig{file=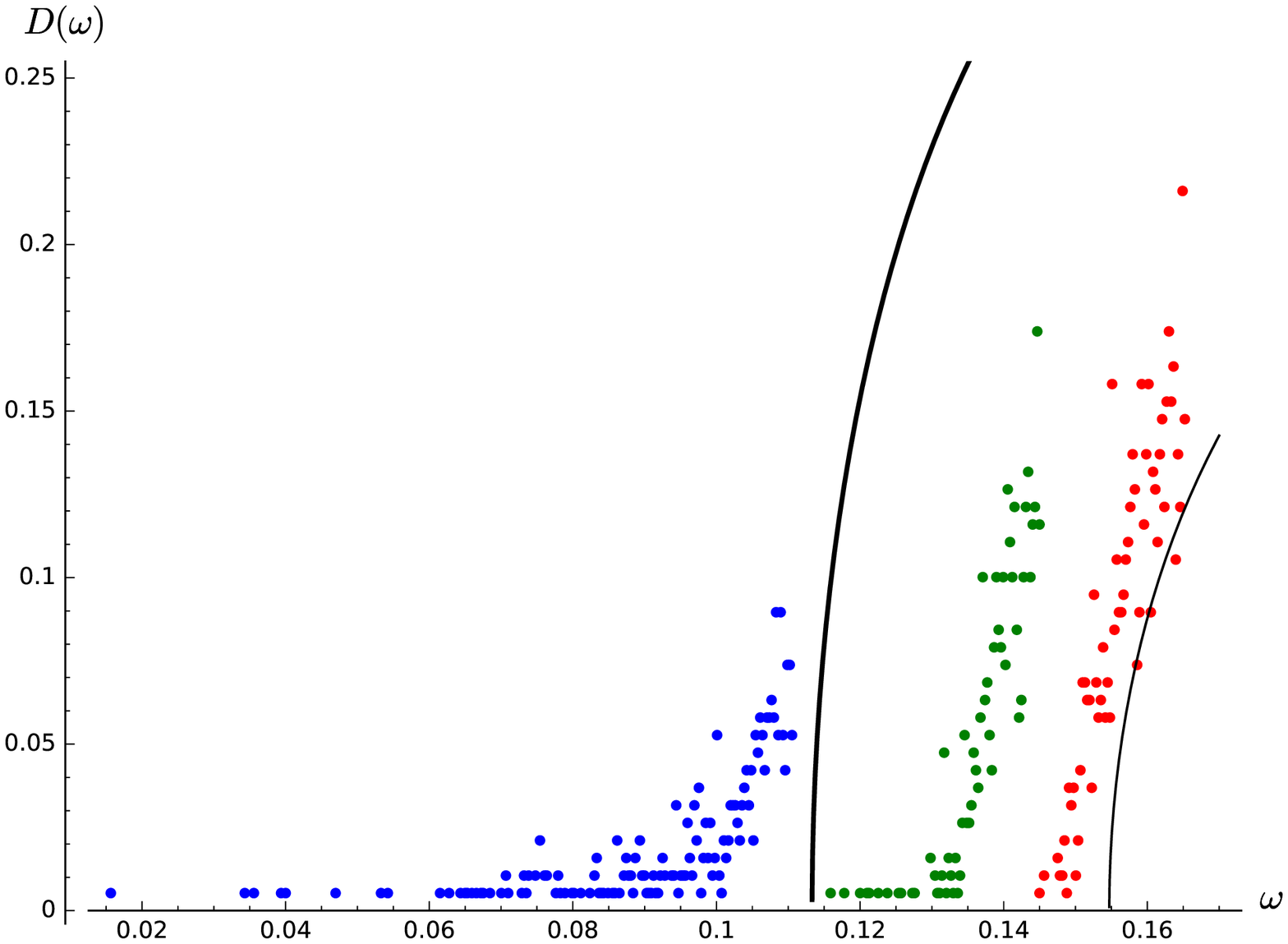, width=4.00cm}
    \caption{The left side figure shows $D(\omega)$ in $d=3$
for $Z=7$ (blue), $Z=8$ (red) and $Z=9$ (gray) for $K_3=0$.
The smooth curves of the corresponding colors are for the MP
DOS.
The center figure  shows the DOS in $d=3$
for $Z=7$ (blue), $Z=8$ (red) and $Z=9$ (gray) for $K_3=0.5$, $K_3=0.47$
and $K_3=0.42$ respectively.
The smooth curves of the corresponding colors are for the SKM DOS, 
whose van Hove peaks are not shown beyond height $0.7$.
In the right hand figure the $Z=7$ DOS is plotted for small $\omega$,
for the case $K_3=0$ (red) and $K_3=0.2$ (green)
and $K_3=0$ (blue points). The points are relative at bins of size
    $\Delta \omega = 10^{-3.5}$.
}
\label{Figdens789}
\end{center}
\end{figure*}

Let us now consider the hyperstatic cases $Z=7,8$ and $9$ in $d=3$.
In the case $K_3=0$ the density of states is fairly
close to the MP DOS with parameter $t=\frac{Z}{d}$;
the DOS' are displayed in the left hand figure in Fig. \ref{Figdens789}.
For $K_3 \le 0.5$ and
as large as managed (i.e. $K_3=0.5$ for $Z=7$, $K_3=0.47$ for
$Z=8$ and $K_3=0.42$ for $Z=9$), the DOS is close, for $\omega < 1.5$,
to the SKM DOS with parameter $t=\frac{Z}{d}$. 
They are plotted in the center hand figure in Fig. \ref{Figdens789}.

The quasi-gap in $D(\omega)$ for $K_3=0$ (estimated in \cite{benetti} 
to go as $\omega^4$ for $\omega\to 0$), becomes a gap for $K_3$ large, as shown in the case
$Z=7$ in the right figure in Fig. \ref{Figdens789}.

\begin{figure*}[h]
\begin{center}
\epsfig{file=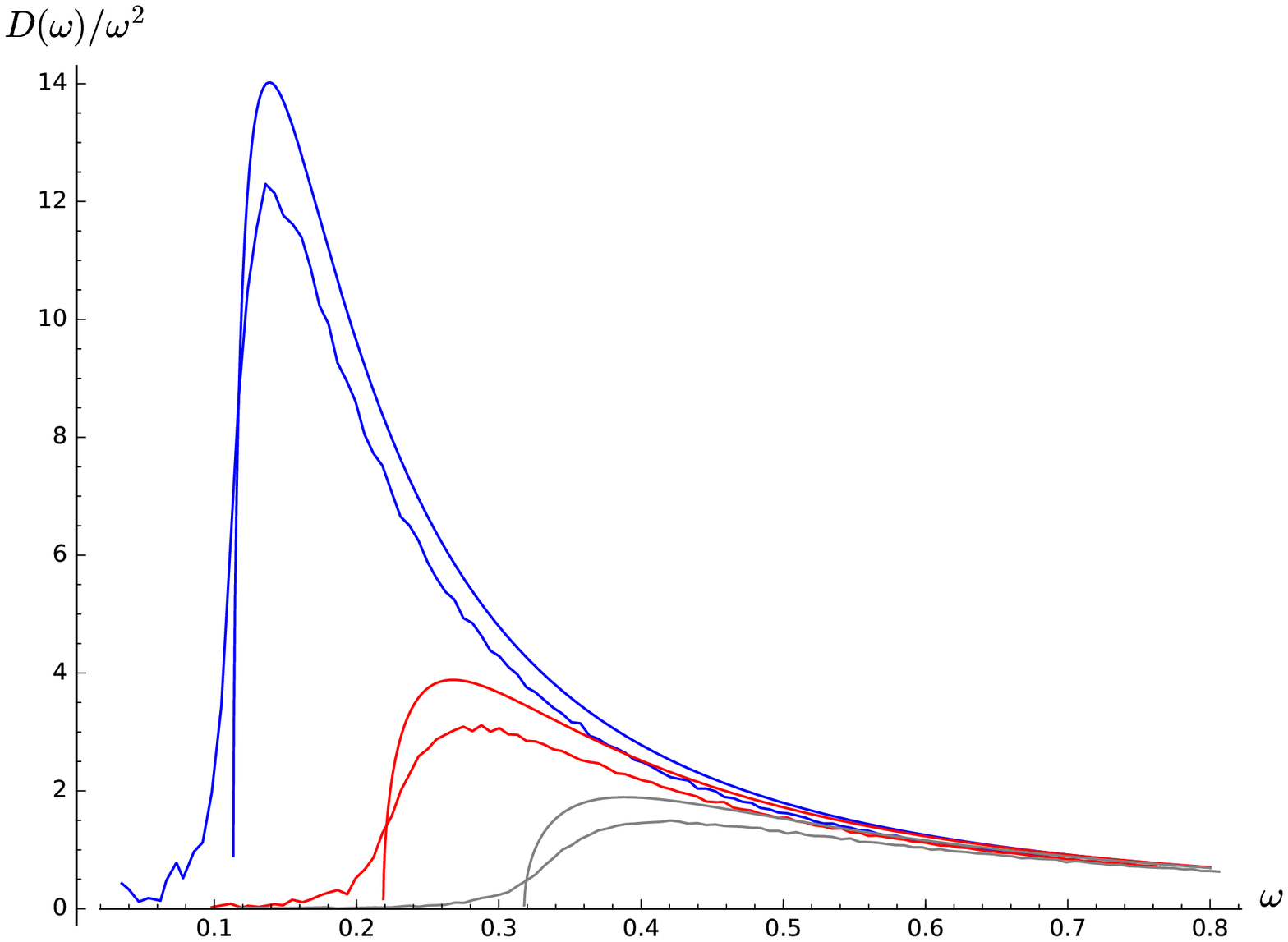, width=7.00cm} \qquad \epsfig{file=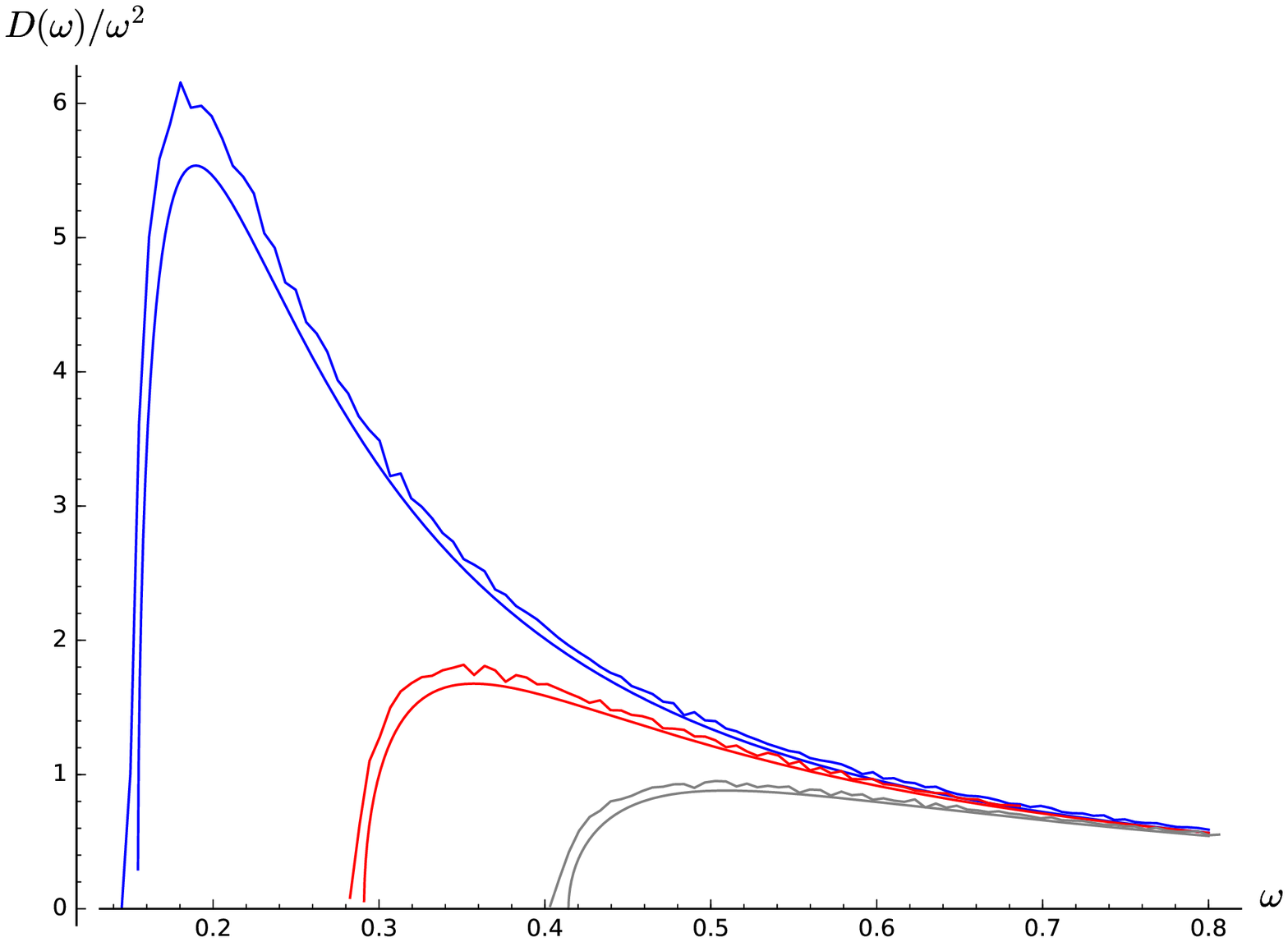, width=7.00cm}
\caption{The left side figure shows the boson peak in $d=3$
for $Z=7$ (blue), $Z=8$ (red) and $Z=9$ (gray) for $K_3=0$.
The smooth curves of the corresponding colors are for the MP DOS.
The right side figure  shows the boson peak in $d=3$
for $Z=7$ (blue), $Z=8$ (red) and $Z=9$ (gray) for $K_3=0.5$, $0.47$
and $0.42$ respectively.
The smooth curves of the corresponding colors are for the SKM DOS.
}
\label{Figbp789}
\end{center}
\end{figure*}

In Fig. \ref{Figbp789} the corresponding boson peaks are shown.

We have seen in a few cases above that in the case of a random $Z$-regular
matrix network and with large $K_3$ cutoff the DOS is close to 
$D_K(\omega; t)$.

As in \cite{benetti}, we consider also a class of models with
random graphs in ${\cal G}_{(z_0,\zeta)}$, i.e. random regular graphs of 
degree $z_0$,  on which Erd\"os-Renyi graphs with average degree $\zeta$
are superimposed. 
This allows to consider real values for $Z=z_0+\zeta$.
We have already examined the $d=1$ ${\cal G}_{(z_0,\zeta)}$ models 
in section III.

With the introduction of a large cutoff $K_d \le 0.5$, the $d=2,3,4$ models
turn out to have a DOS closer to that of the $d=1$ model with the same $t$, 
and hence to the SKM DOS $D_K(\omega; t)$, than to the MP DOS.

\begin{figure*}[h]
\begin{center}
\epsfig{file=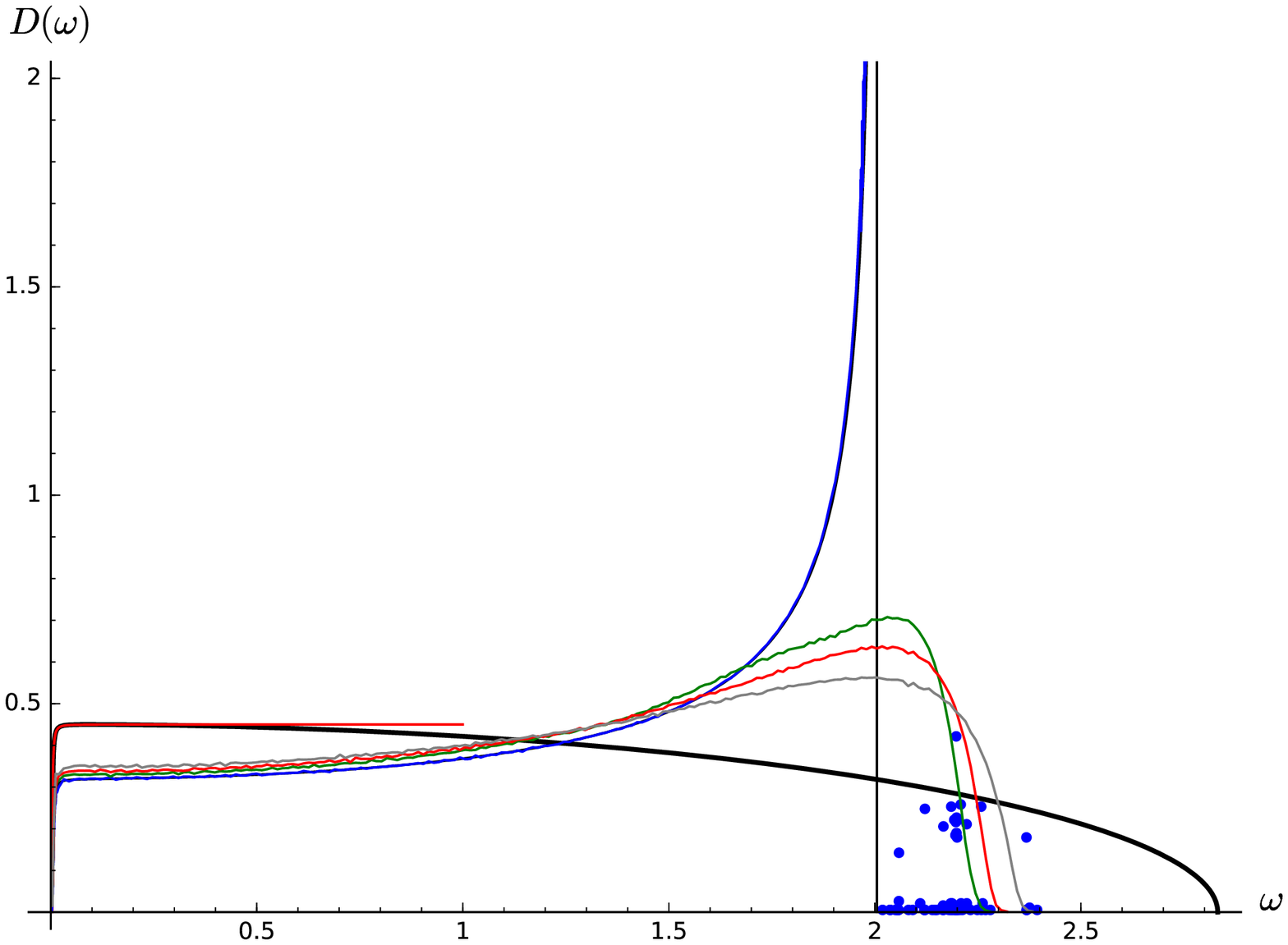, width=5.00cm} \qquad \epsfig{file=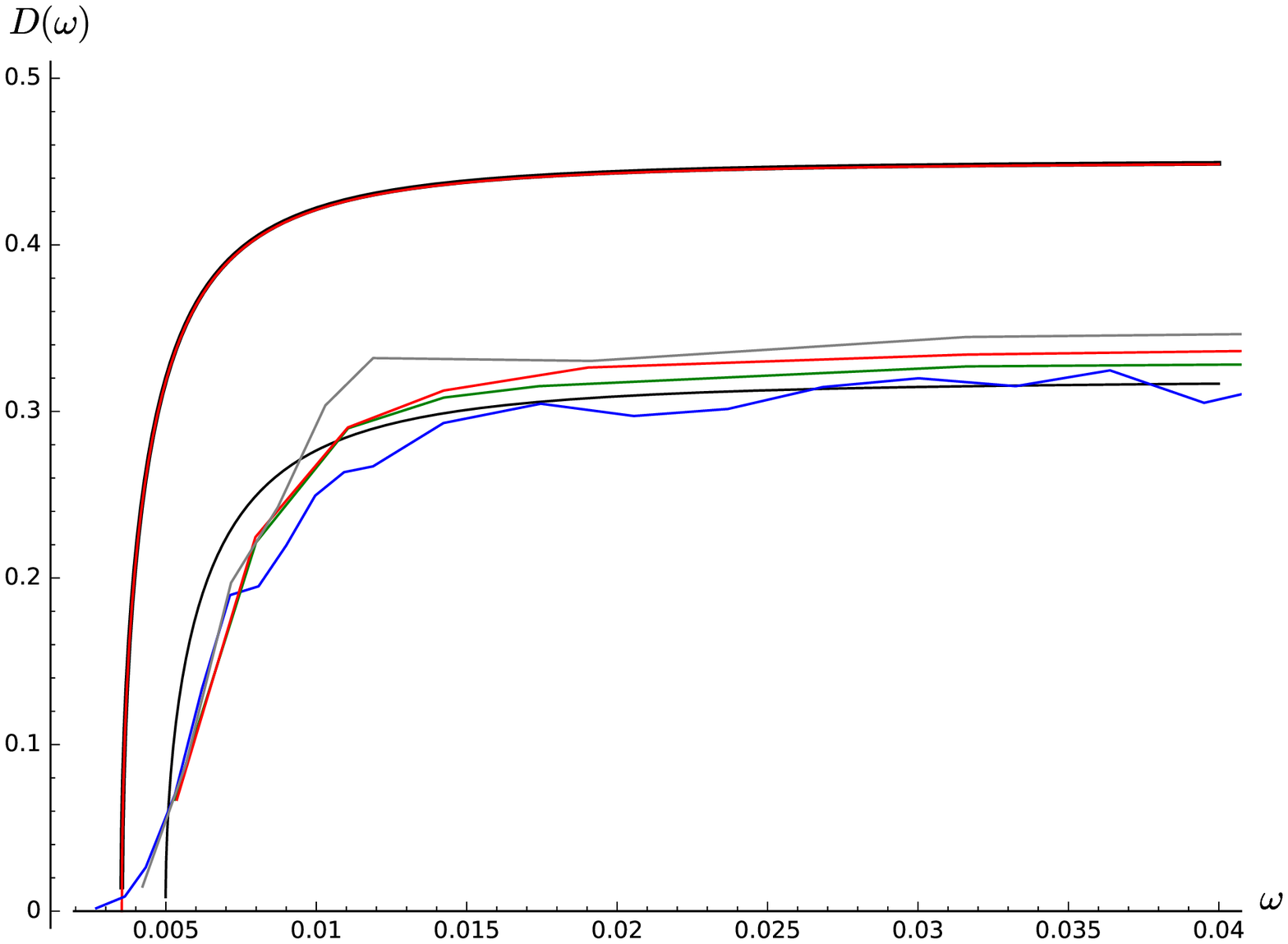, width=5.00cm} \qquad \epsfig{file=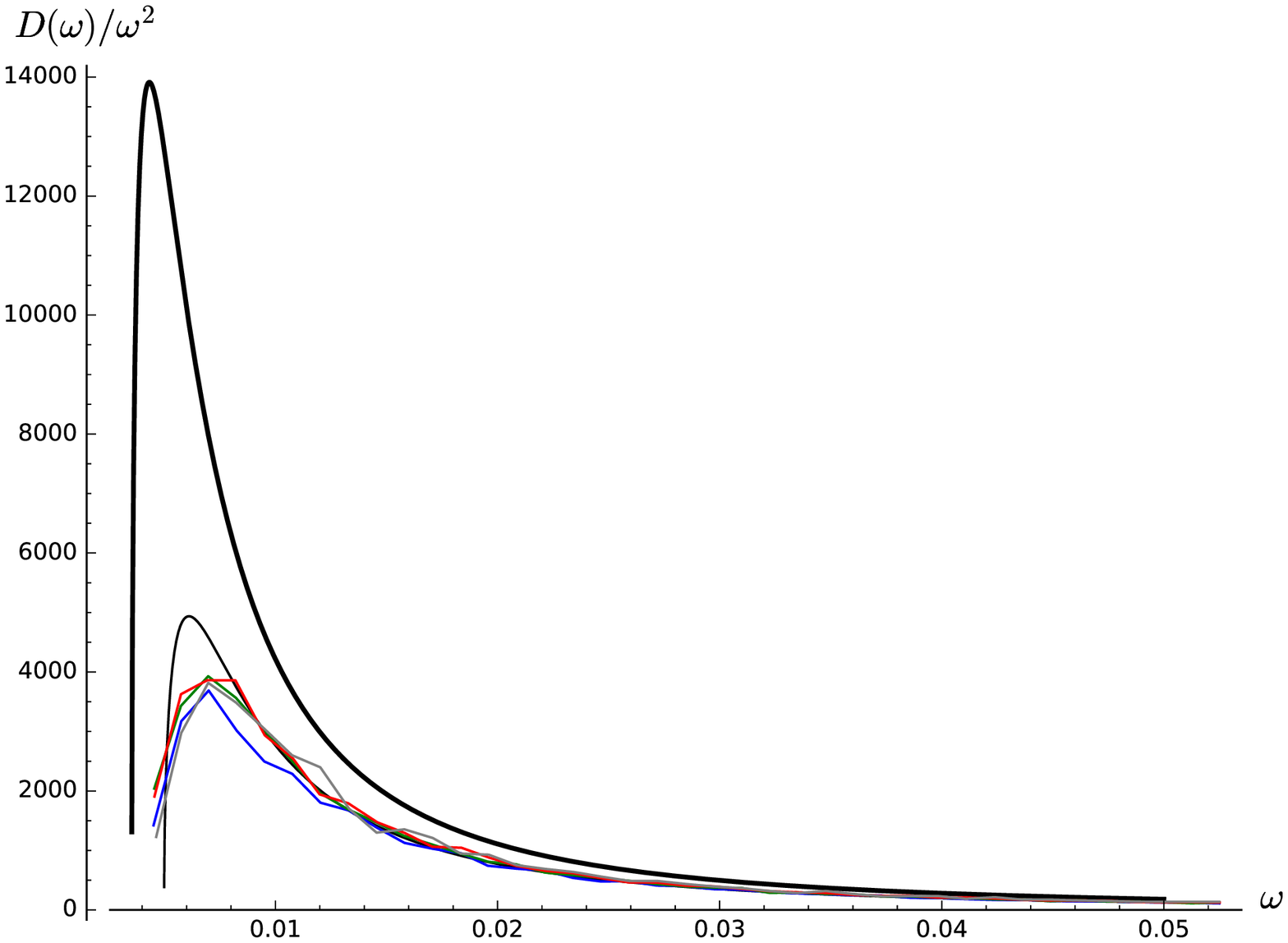, width=5.00cm}
\caption{The left side figure shows the DOS for
    $t=2.01$ for $d=1$ (blue), $d=2$ with $K_2=0.4$ (green), $d=3$ with
$K_3=0.5$ (red) and $d=4$ with $K_4=0.5$ (gray).
In the case $d=1$ the HF part is not a smooth curve, so
    points are given, representing bins of width $\Delta \omega = 10^{-3.5}$.
A point in $(2.1976(3), 11.8)$ is not represented in the figure.
The center figure shows $D(\omega)$ for small $\omega$ for the same cases.
The thin black curve is the SKM DOS, the thick
black curve is the MP DOS, the red curve close to it is the EMA DOS in Eq.(\ref{ema}).
The right figure shows $\frac{D(\omega)}{\omega^2}$ for the same cases.
}
\label{Fig201}
\end{center}
\end{figure*}

In Fig. \ref{Fig201} are given the results of simulations in the case
$t=\frac{Z}{d} =2.01$. For the case $d=1$ it is the case
${\cal G}_{(2,0.01)}$, which we have discussed in section III.
Here we plot its DOS; for $\omega < 1.5$, it is close to the SKM DOS and to the DOS
of the models with large $K_d$ in $d=2$, $3$ and $4$ with the same $t$.
The highest values found for the cutoff are $K_2=0.4$, $K_3=K_4=0.5$,
using $1500$ nodes in $d=2$, $1000$ nodes in $d=3$ and $750$ nodes in
$d=4$; for each case, we make the average on $200$ random graphs.

\begin{figure*}[h]
\begin{center}
\epsfig{file=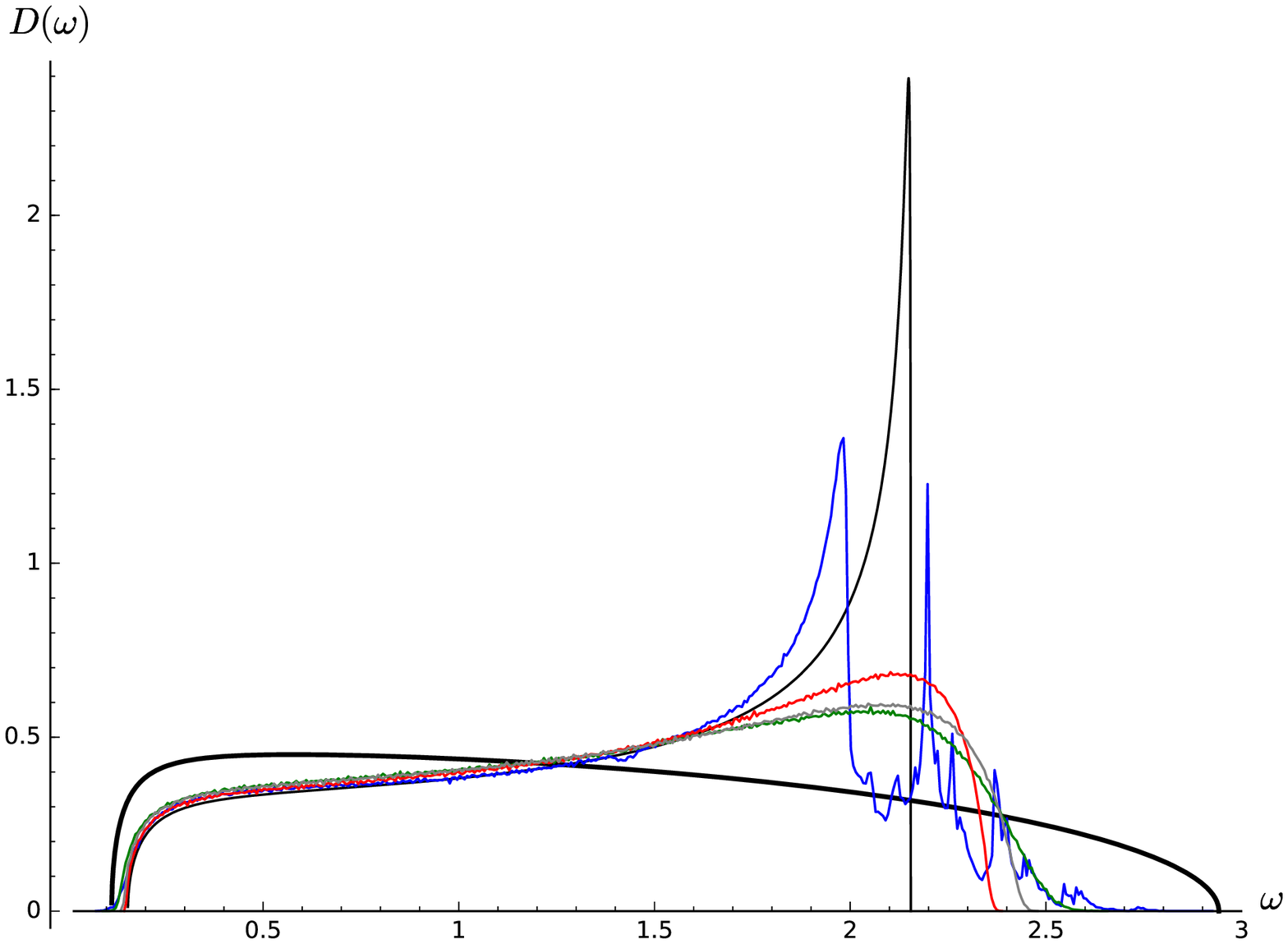, width=5.00cm} \qquad \epsfig{file=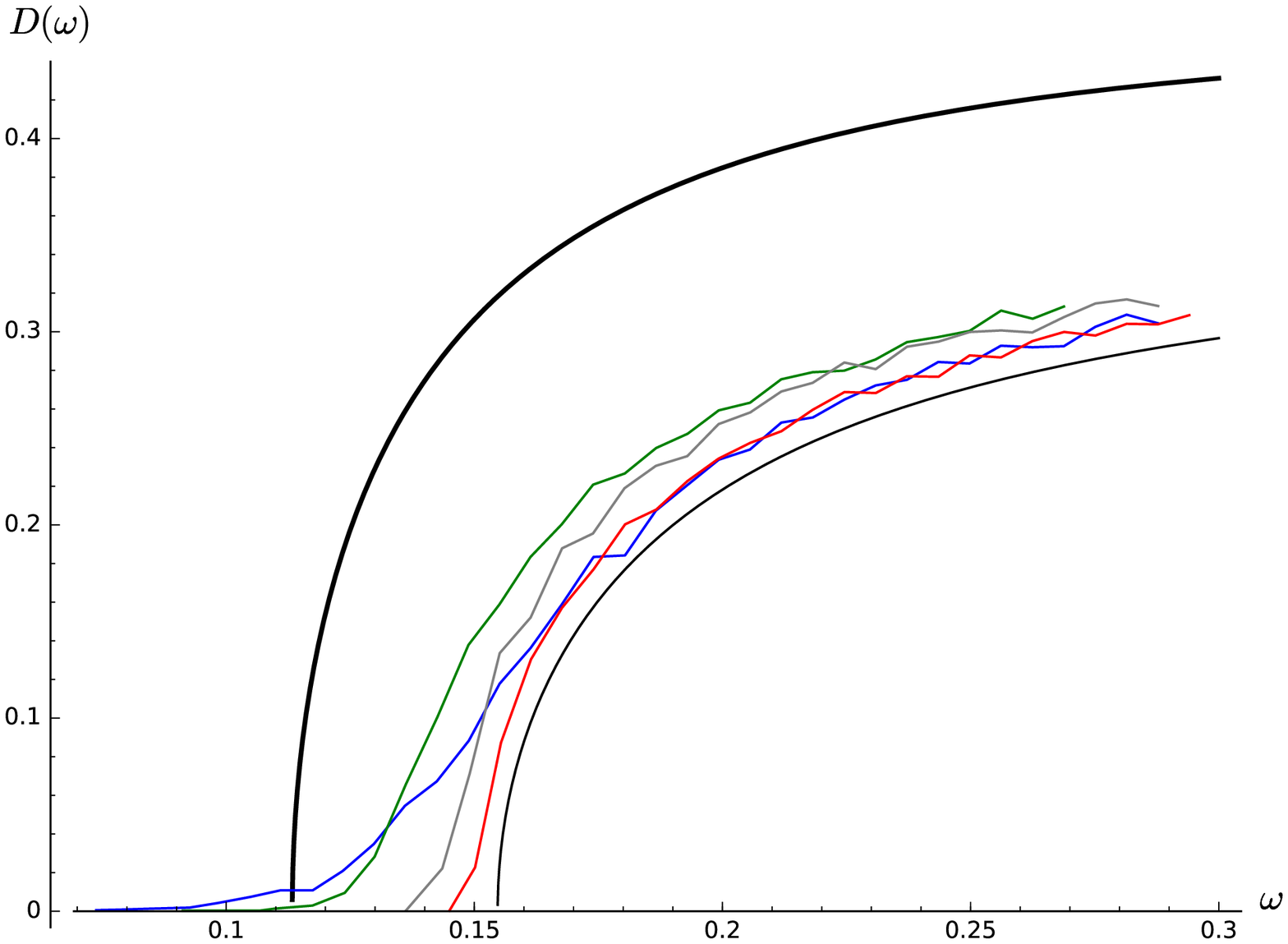, width=5.00cm} \qquad \epsfig{file=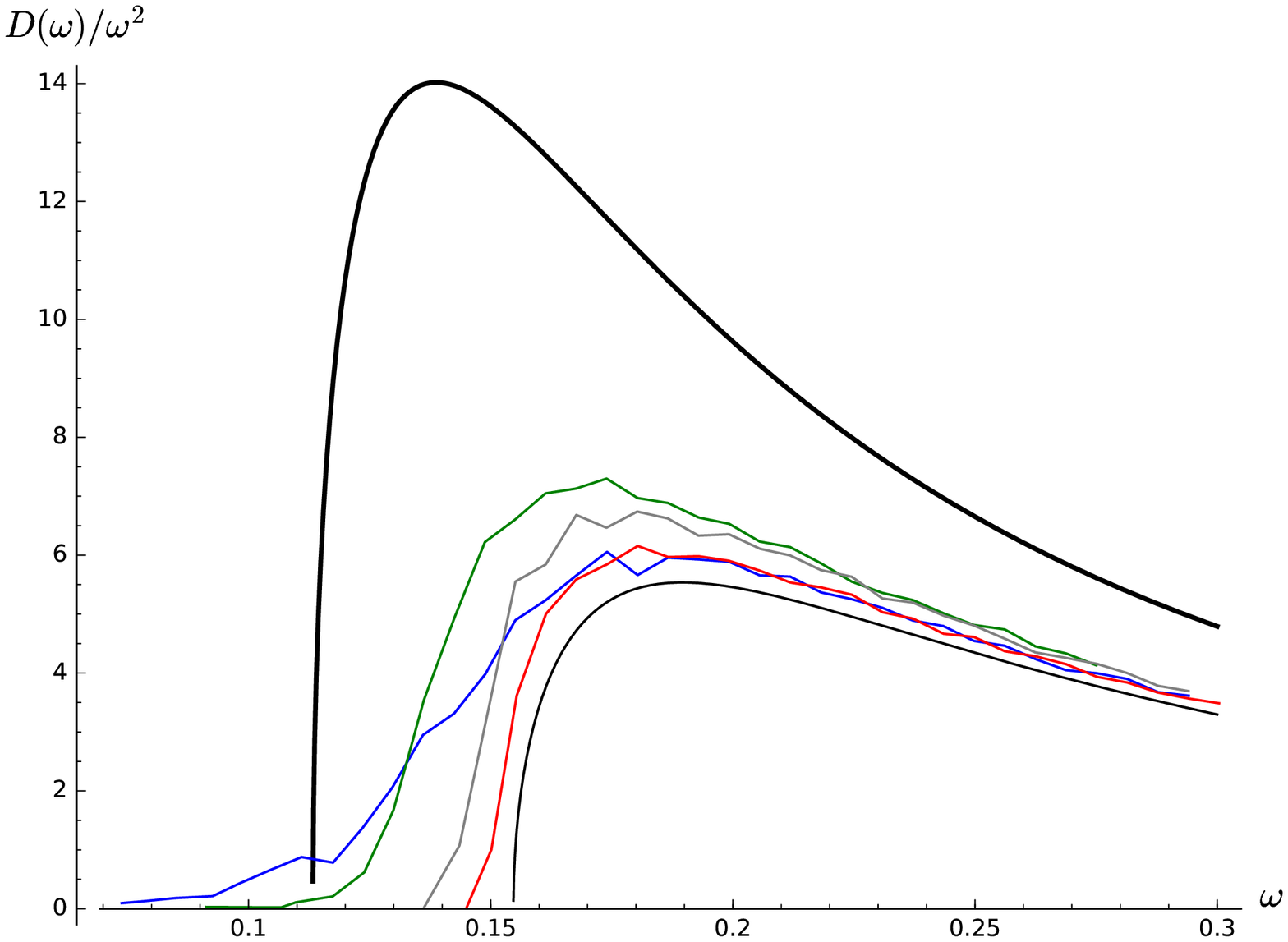, width=5.00cm}
\caption{The left side figure shows the density of states for
    $t=7/3$ for $d=1$ (blue), $d=2$ with $K_2=0.16$ (green), 
$d=3$ with $K_3=0.5$ (red) and $d=4$ with $K_4=0.5$ (gray).
The center figure shows $D(\omega)$ for small $\omega$ for the same cases;
the right figure shows $\frac{D(\omega)}{\omega^2}$ for the same cases.
The thin black curve is the SKM DOS, the thick
black curve is the MP DOS.
}
\label{Fig233}
\end{center}
\end{figure*}

In Fig. \ref{Fig233} are given the results of simulations in the case
$t = 7/3$. As in the previous case, the densities of states in different 
dimensions, for $K_d$ large, are close to the SKM DOS, for $\omega < 1.5$.

For $t=3$ the DOS for $d=2,3$ and $4$ are close, for $\omega < 1.5$, to $D_K(\omega; 3)$ 
(see the center figure in Fig.\ref{Figdens789} for the case $d=3$), which is
the DOS of the $d=1$ ${\cal{G}}_{(3,0)}$ model.

Halving the number of nodes and making the average on $400$ graphs
we get similar results for the DOS. The maximum values of $K_d$ change
slowly with $N$; in $d=3$ they are less than $5\%$ higher halving $N$.
In $d=2$ the change is greater far form the isostatic point;
for $z_0=2$ and $\zeta=\frac{1}{3}$ the maximum value of $K_2$, halving $N$,
is $13\%$ higher. Therefore it can be expected that increasing $N$ it
decreases further.

\begin{figure*}[!htb]
\begin{center}
\epsfig{file=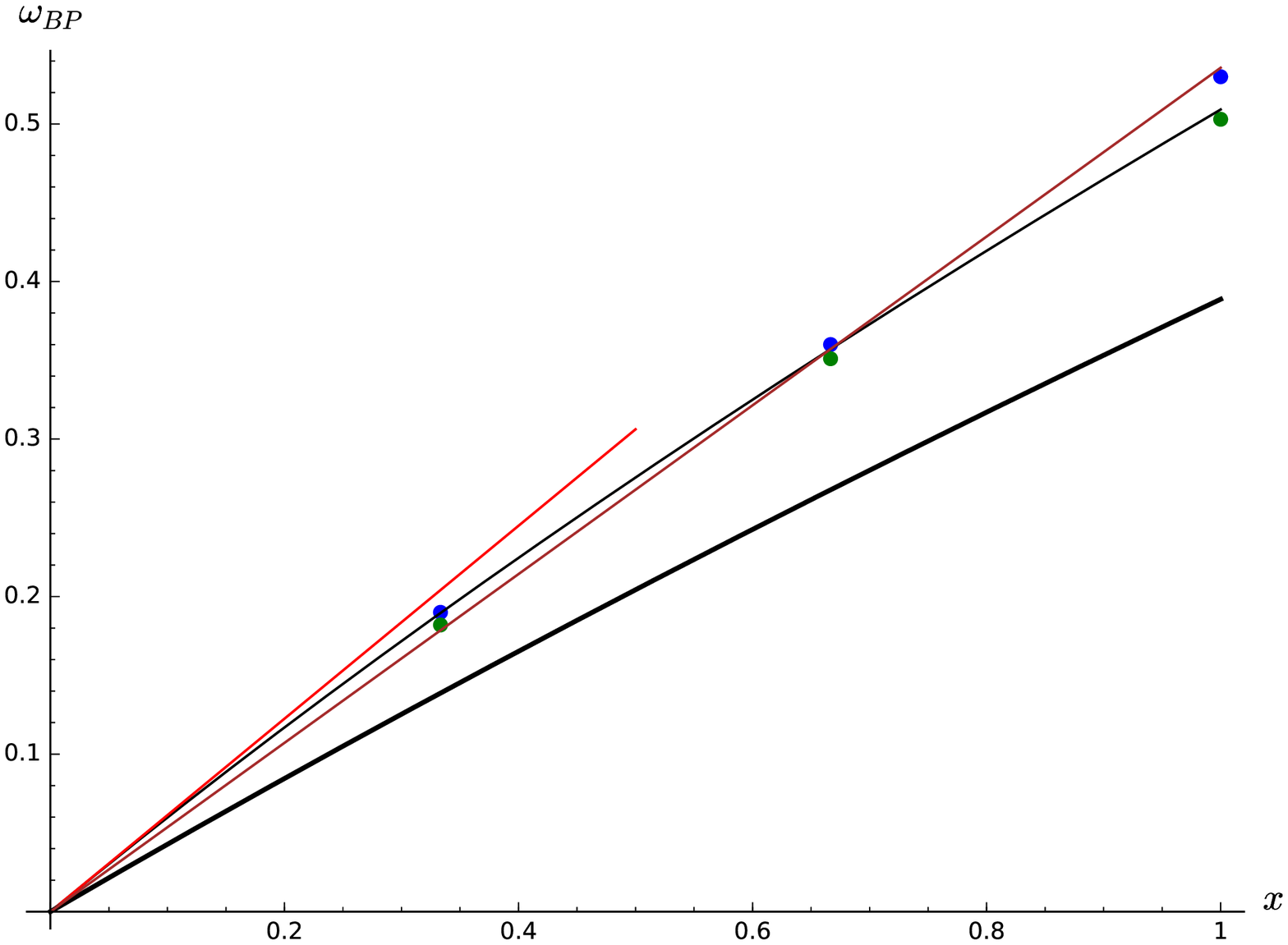, width=7.00cm}\qquad \epsfig{file=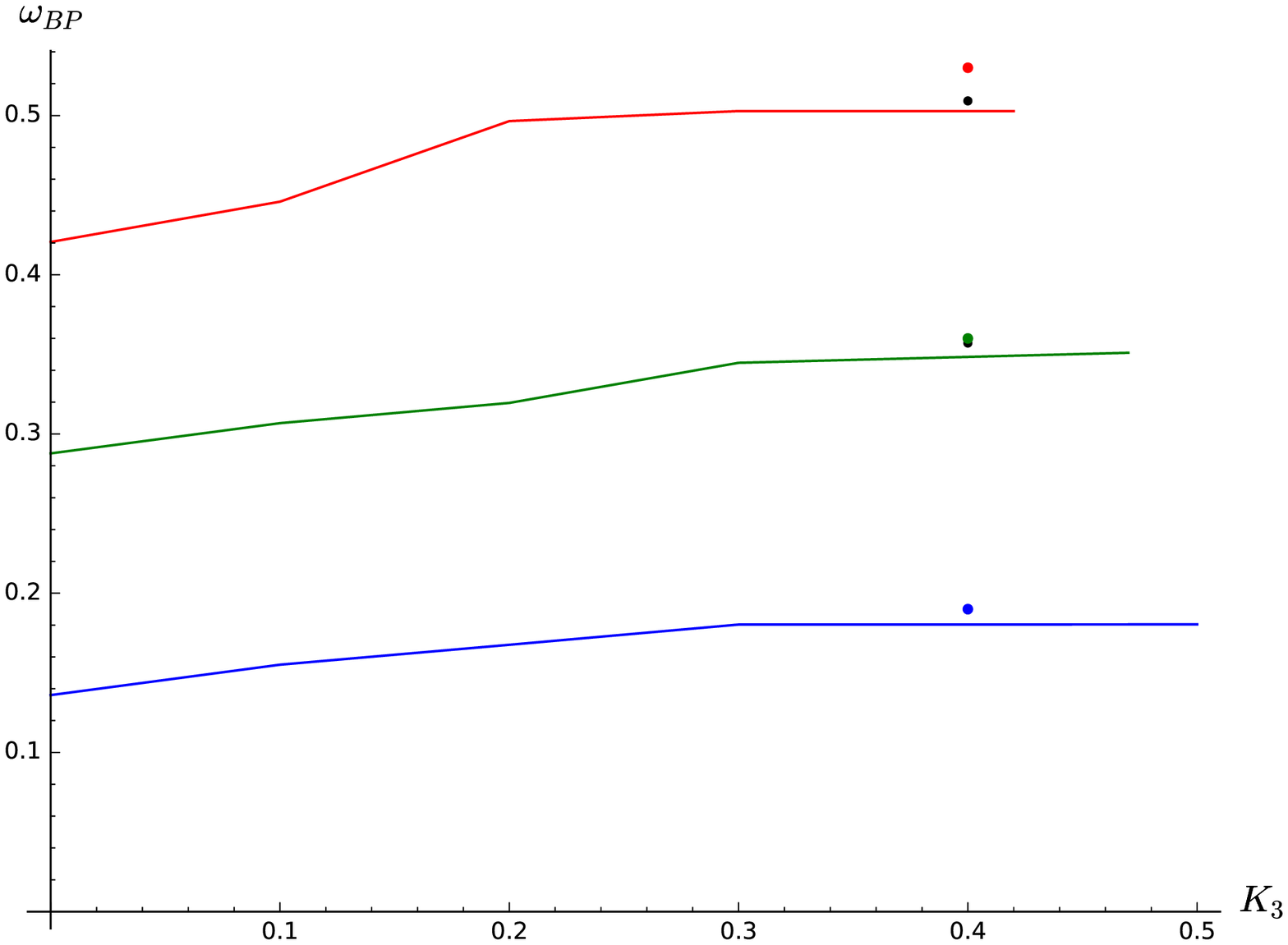, width=7.00cm}
\caption{In the left hand figure,
the boson peak $\omega_{BP}$ (thin black line) is plotted
against $x=t-2$ and its tangent
in $x=0$ (red line); $\omega_{BP}$ is computed as the maximum of 
$\frac{D_K(\omega; t)}{\omega^2}$.
The blue dots are $\omega_{BP}$ in molecular dynamics simulations in 
\cite{MZ} for $Z=7,8,9$ in $d=3$;
the brown line is a linear fit for these points.
The green dots are $\omega_{BP}$ in simulations with largest $K_3$.
The boson peak for the MP DOS is drawn as a thick black line.
In the right hand figure, there is a plot of the boson peak against $K_3$;
the case $Z=7$ is the line in blue, the case $Z=8$ is the green line,
the case $Z=9$ the red line. With the same colors a point indicates
    the values found in molecular simulations \cite{MZ}.
The black points are the corresponding values from the SKM DOS.
}
\label{Figpvbp}
\end{center}
\end{figure*}

The curve of the maxima of 
$\frac{D_K(\omega; t)}{\omega^2}$ is given by
\begin{equation}
\omega^6 - \frac{10}{3} t \omega^4 + \frac{1}{3}(13t^2-20t+20) \omega^2 - 8t 
    +8t^2 - 2t^3 = 0
\end{equation}
The slope of this curve at the isostatic point $t=2$ is $\frac{\sqrt{6}}{4}$; 
see the left hand figure in Fig. \ref{Figpvbp}.
In it are reported the boson peak frequencies in $d=3$, $Z=7$, $8$ and $9$
found in molecular dynamics simulations in \cite{MZ}.
For $Z=7$ and $8$ these values agree well with the SKM curve,
while for $Z=9$ the difference is $4\%$.

In the right hand figure in Fig. \ref{Figpvbp}, $\omega_{BP}$ is plotted
against $K_3$.
There is little difference in $\omega_{BP}$ for $K_3=0.3$ and larger values
of it. 

In the isostatic case we saw that for $K_3 \simeq 0.3$ the height of
the plateau around $\omega=0$ fits with the molecular simulation results
in \cite{ohern}.

Therefore overall $K_3 \simeq 0.3$ is a good fit in all these cases.

\clearpage

\section{Conclusion}
We find that the shifted Kesten-McKay density of states with parameter 
$Z=z_0+\zeta$ is a mean
field solution in a class of random matrix models, based on
(random) regular graphs of degree $z_0$ to which an Erd\"os-Renyi graph
with small average degree $\zeta$ is superimposed.
We conjecture that for $\zeta \to 0$ the cumulative function
of the density of states of this model tends uniformly to the one of the
 mean-field solution,
in an interval $[0, \omega_0]$, with $\omega_0 < \sqrt{z_0-1} + 1$.
The $z_0=2$ case is the $k=1$ Newman-Watts small-world model.

We introduce a random block matrix model for soft spheres near the 
jamming point; it differs from the random elastic network model introduced
in \cite{pari1, benetti}, due to the introduction of an angular cutoff
$K_d$ between contact versors, modeling the excluded volume due to sphere
repulsion in an equilibrium configuration.

Making simulations in $2 \le d \le 4$ with the same class of random graphs 
as above, we find that the density of states
in the case of large $K_d$ is close to the shifted Kesten-McKay density 
of states with parameter $t=\frac{Z}{d}$, 
while in the elastic network case $K_d=0$ it is close to
the Marchenko-Pastur density of states with parameter $t=\frac{Z}{d}$.

The  shifted Kesten-McKay density 
of states gives in $d=3$, $Z=7$ and $8$ boson peak frequencies in 
good agreement with molecular dynamics simulations with soft 
spheres \cite{MZ}.
With respect to the random elastic network model, the large $K_d$ model
has fewer low-frequency
modes; in the isostatic case it has a plateau around $\omega=0$,
whose height is for $K_3 \simeq 0.3$ 
in agreement with molecular dynamics simulations with soft spheres 
\cite{ohern}.

The density of states found in simulations, in the case of large $K_d$,
has in the hyperstatic case a gap around $\omega=0$, 
a bit smaller than the one of the shifted Kesten-McKay density 
of states; therefore $D(\omega)$ does not go
as $\omega^4$ for $\omega \to 0$. In the elastic network model $K_d=0$
there is instead a quasi-gap, for which the behavior $D(\omega) \sim \omega^4$
has been found in \cite{benetti}.
\\

I thank Gianni Cicuta and Alessio Zaccone for discussions.
\\


\end{document}